\documentclass[sigplan,nonacm]{acmart}

\usepackage[inline]{enumitem}
\usepackage{makecell}
\usepackage{multirow}
\usepackage{subcaption}

\begin{document}

\title{Orchestrating multi-level magic state distillation: a dynamic pipeline architecture}

\author{Junshi Wang}
\affiliation{
	\institution{University of Cambridge}
	\city{Cambridge}
	\country{United Kingdom}
}
\email{jw2452@cam.ac.uk}

\author{Prakash Murali}
\affiliation{
	\institution{University of Cambridge}
	\city{Cambridge}
	\country{United Kingdom}
}
\email{pm830@cam.ac.uk}

\begin{abstract}

Practical quantum computation requires high-fidelity instruction executions on qubits. Among them, Clifford instructions are relatively easy to perform, while non-Clifford instructions require the use of magic states. This makes magic state distillation a central procedure in fault-tolerant quantum computing. A magic state distillation factory consumes many low-fidelity input magic states and produces fewer, higher-fidelity states.
To reach high fidelities, multiple distillation factories are typically chained together into a multi-level pipeline, consuming significant quantum computational resources. 
Our work optimizes the resource usage of distillation pipelines by introducing a novel dynamic pipeline architecture. Observing that distillation pipelines consume magic states in a burst-then-steady pattern, we develop dynamic factory scheduling and resource allocation techniques that go beyond existing static pipeline organizations.
Dynamic pipelines reduce the qubit cost by 16\%--70\% for large-scale quantum applications and achieve average reductions of 26\%--37\% in qubit--time volume on generated distillation benchmarks compared to state-of-the-art static architectures.  
By significantly reducing the resource overhead of this building block, our work accelerates progress towards the practical realization of fault-tolerant quantum computers.

\end{abstract}

\maketitle

\section{Introduction}
\label{sec:introduction}

Quantum computing is a computational paradigm that leverages quantum mechanics to solve problems that are intractable on classical computers.
Current noisy intermediate-scale quantum (NISQ) devices are susceptible to physical noise~\cite{Preskill2018nisq}, resulting in high error rates that limit the size of the problems that these devices can solve.
To achieve practical quantum advantage, we must tackle large problems that scale beyond classical computational limits~\cite{hoefler2023disentangling}. This requires fault-tolerant quantum computing (FTQC) where quantum error correction (QEC) is used to protect quantum information by redundantly encoding quantum information across multiple physical qubits. This results in \emph{logical} qubits which have improved error rates compared to their physical qubits~\cite{Mermin_2007}.

Universal quantum computation on logical qubits requires both Clifford and non-Clifford gates~\cite{gottesman1998heisenberg}, but typical QEC codes natively support only Clifford gates~\cite{fowler2013surface}. To support non-Clifford gates, \emph{magic state injection} is used. That is, a magic state is prepared using some procedure and injected into the logical qubit where the non-Clifford operations is desired. The quality of the injected states directly impacts the logical error rate, making the preparation of high-quality magic states an important component of a quantum computer.
To prepare high-quality magic states, \emph{magic state distillation}~\cite{bravyi2005universal} is the standard method. It converts multiple low-fidelity input magic states into fewer, high-fidelity ones. 
For example, the Reed-Muller protocol~\cite{bravyi2005universal} takes $15$ input noisy magic states to produce one high-quality magic state, reducing error rates by a cubic factor. Since practical applications will require magic states in fidelities below $10^{-10}$, one round of distillation typically does not suffice~\cite{beverland2022assessing}. Distillation is usually performed in a multi-level fashion, where each level includes a set of distillation circuits (factories), ultimately producing very high fidelity magic states~\cite{jones2013multilevel}.
This process, however, is costly: some applications devote up to 95\% of their total qubits to distillation~\cite{beverland2022assessing}.
Optimizing this multi-level distillation pipeline is therefore the focus of our work.

\begin{figure*}[htbp]
\centering
\includegraphics[width=0.93\linewidth]{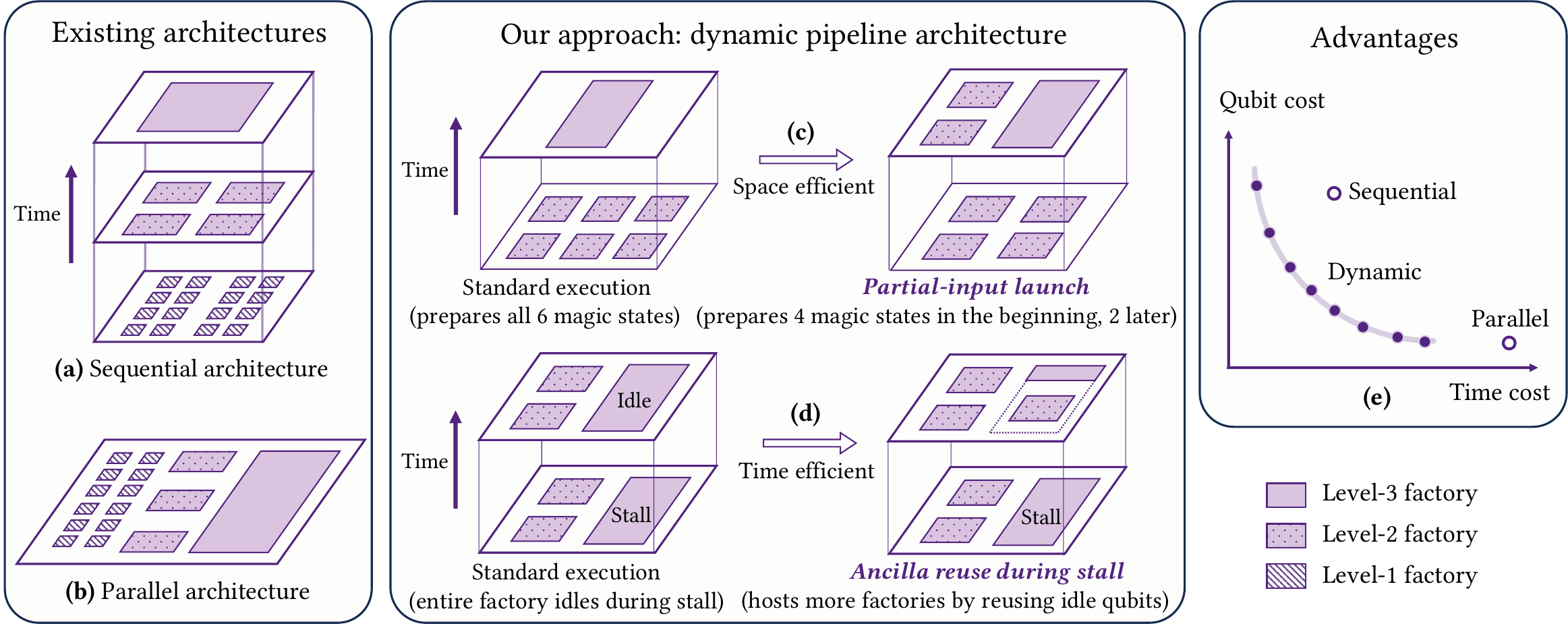}
\caption{Existing distillation pipeline architectures and our dynamic architecture.
    Each horizontal plane represents a snapshot of the running factories at a particular time step.
    Factories at higher levels use logical qubits with larger code distances, leading to larger patches.
    \textbf{(a)} Each distillation level executes in sequence. Many low-level factories can run in parallel, with only a few high-level factories can be packed within the available qubits.
    \textbf{(b)} All levels are allocated dedicated qubit regions and execute in parallel.
    \textbf{(c)} Partial-input launch allows launching high-level distillation before all input magic states are ready, reducing the number of active low-level factories and saving qubits.
    \textbf{(d)} When high-level factories stall due to insufficient magic-state input, some regions of the idling high-level factory can be reused to host more low-level factories, making subsequent distillation faster.
    \textbf{(e)} Our architecture not only reduces both qubit and time usage, but also exposes the full space--time trade-off space.
}
\label{fig:1:overview}
\end{figure*}

Prior works on distillation pipelines adopt fixed architectures, where the pipeline structure is statically defined and does not adapt to runtime conditions. There are two types of fixed pipelines: sequential and parallel.
In the sequential pipeline~\cite{beverland2022assessing}, distillation levels are executed in a strict sequence on the same set of qubits (Fig.~\ref{fig:1:overview}(a)). The scheduling order and resource assignments are predetermined and unchanging. 
In the parallel pipeline~\cite{silva2024optimizing}, all levels run concurrently in dedicated qubit regions with one region feeding states to the next region (Fig.~\ref{fig:1:overview}(b)). 
The number of factories at each level is fixed in advance to match production and consumption rates. At face value, the sequential architecture uses fewer qubits but takes more time, while the parallel architecture achieves lower execution time at the expense of more qubits. However, our work shows that 
both designs suffer from resource inefficiencies due to their rigid pipeline structures, leaving many qubits underutilized.  Importantly, we also observe that these works assume a factory can begin only when previous levels of the pipeline are completed. 

Our work proposes a novel dynamic magic state distillation architecture, shown in Figure~\ref{fig:1:overview}. Our work rests on the observation that magic-state factories 
exhibit a burst-then-steady consumption pattern: it consumes a burst of magic states at the beginning and then consumes them gradually.
This enables a supply-driven perspective to design the pipeline, where magic states are only required to be supplied on time to meet the consumption pattern, rather than preparing all required magic states before execution.
Based on this observation, we design a dynamic distillation pipeline architecture that allows the pipeline structure to evolve dynamically based on the magic-state availability from currently completed distillation levels and available compute resources.

To enable dynamism, we use two strategies. First, \emph{partial input launch} allows high-level factories to start execution as soon as some inputs are available, with the remaining inputs supplied gradually (Fig.~\ref{fig:1:overview}(c)). Second, when a high-level factory stalls due to insufficient inputs, we introduce \emph{ancilla qubit reuse during stalls} to repurpose qubits to host additional low-level factories and  accelerate the production of magic states (Fig.~\ref{fig:1:overview}(d)). 

We tackle the challenge of constructing an efficient dynamic pipeline with these two strategies. We break the problem of multi-level distillation into a series of independent subproblems, each involving only a two-level pipeline.
To optimize two-level pipelines, we introduce
\begin{enumerate*}[label=(\arabic*)]
    \item a dynamic scheduler, which addresses the temporal dimension by determining when factories should be launched, and
    \item a resource allocator, which addresses the spatial dimension by selecting the appropriate types of low-level factories and allocating qubits to these factories and the buffer.
\end{enumerate*}

We implemented our architecture in simulation and compared its resource usage to state-of-the-art static designs.
On large-scale real-world quantum applications, our architecture reduces total qubit requirements by up to 70\% (Heisenberg model) and 30\% (Ising model) compared to sequential~\cite{beverland2022assessing} and parallel~\cite{silva2024optimizing} baselines, respectively, with most cases showing at least a 25\% reduction.
On generated distillation benchmarks, it achieves up to 65\% and 31\% reductions in qubit--time volume (with average reductions of 37\% and 26\%) compared to the baselines.
Our key contributions are:

\begin{itemize}
\item Our work is the first to propose the dynamic magic distillation pipelines, overcoming key limitations of existing works~\cite{beverland2022assessing,silva2024optimizing}. 
\item Dynamic pipelines offer significant resource improvements over static pipelines. They can be implemented with additional software control and do not need any fundamental changes to quantum error correction or qubit design. This makes them an attractive technique for future FTQC system designs. 
\item Our work exposes new trade-offs between distillation space (qubits) and time (Fig.~\ref{fig:1:overview}(e)).
While existing architectures each yields only a single distillation pipeline configuration (a fixed point on the space--time diagram), our method produces a spectrum of configurations that form the Pareto front. This allows quantum architects to select the configurations best suited to their hardware capabilities and application requirements.

\end{itemize}

\section{Background}
\label{sec:background}

\subsection{Fault tolerant quantum computing}
\paragraph{Surface code.}

The surface code is a leading QEC scheme due to its practical hardware requirements~\cite{fowler2013surface, fowler2018low} and has been prototyped experimentally~\cite{google2023suppressing}.
A logical qubit is encoded in a $d \times d$ patch of physical qubits, where the \emph{code distance} $d$ determines error suppression: larger $d$ offers stronger protection but requires more qubits and longer runtime.

Error correction relies on repeated stabilizer (parity) measurements using ancilla qubits.
A round of stabilizer measurement has hardware-determined constant duration
\begin{equation}
    T_{\text{stab}} = 6 T_{\text{2q}} + T_{\text{meas}}\,,
    \label{eq:2:stab}
\end{equation}
where $T_{\text{2q}}$ and $T_{\text{meas}}$ are the durations of two-qubit gates and measurement, respectively.
One logical cycle includes $d$ consecutive stabilizer measurements, so it takes $d\cdot T_{\text{stab}}$ time.
The physical-to-logical error suppression per cycle can be approximated by
\begin{equation}
p_L \approx 0.03 \left( \frac{p}{0.01} \right)^{\frac{d+1}{2}}\,,
\label{eq:2:surface-error}
\end{equation}
where $p$ is the physical error rate and the constants are numerically determined~\cite{fowler2013surface,beverland2022assessing}.
We assume the surface code as our underlying code, following existing studies on distillation pipelines and resource estimation~\cite{beverland2022assessing, gidney2021factor, litinski2019game, litinski2019magic}, though our ideas are widely applicable across QEC codes. Our work focuses on logical qubits rather than physical qubits.

\paragraph{Logical operations.}
The standard method for implementing Clifford logical operations (e.g., CNOT and H) in the surface code is lattice surgery~\cite{horsman2012surface}.
Quantum circuits are compiled into sequences of multi-Pauli measurements~\cite{silva2024multi,litinski2019game}, which perform Clifford operations by merging and splitting the involved logical qubit patches.
These patches can also be moved by deforming and relocating to the target position, requiring only one logical cycle regardless of the movement distance~\cite{litinski2019game}.
For non-Clifford operations (e.g., $T$), lattice surgery is combined with magic state injection, which consumes qubits that have been pre-prepared in the special state $|m\rangle = \tfrac{1}{\sqrt{2}}\left(|0\rangle + e^{i\pi/4}|1\rangle\right)$, known as the \emph{magic state}~\cite{bravyi2005universal,horsman2012surface}.

\subsection{Magic state distillation}

\begin{figure}[tbp]
    \centering
    \includegraphics[width=\columnwidth]{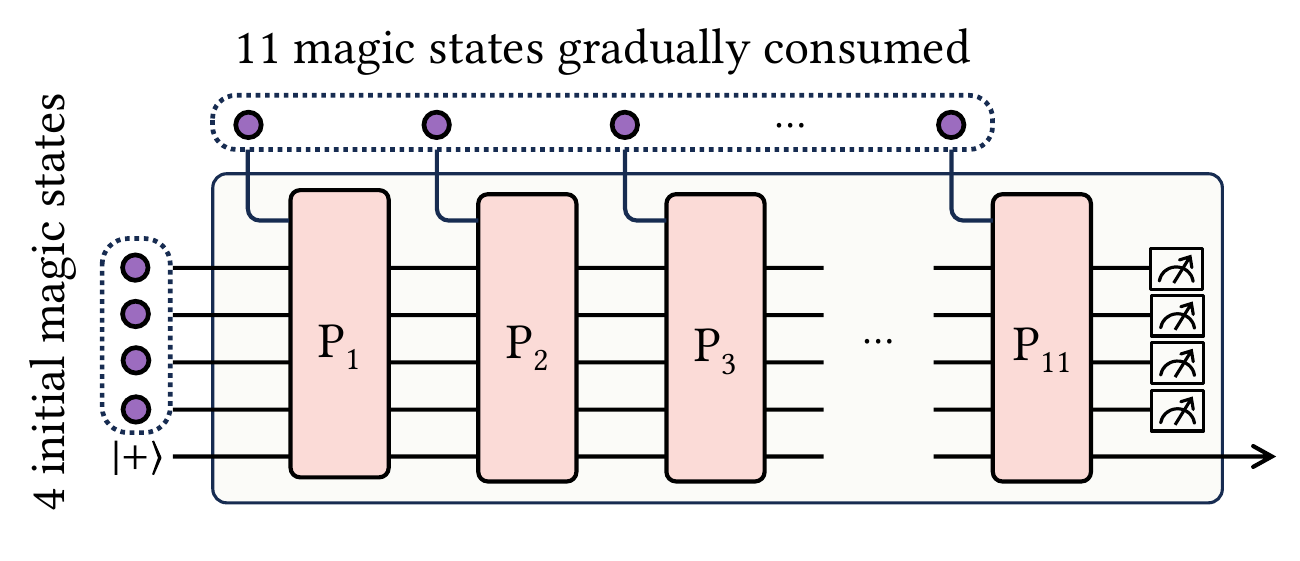}
    \caption{A 15-to-1 distillation factory.
    It contains 11 non-Clifford rotations $P_1,\cdots,P_{11}$ (defined in Ref.~\protect{\cite{litinski2019game}}), each consuming one input magic state.
    A total of 15 input states are consumed in a burst-then-steady pattern; 4 are consumed initially and 11 gradually.
    A higher-fidelity output is produced in the end if all measurements yield $+1$; otherwise the protocol fails and discards the output.
    }
    \label{fig:2:factory}
\end{figure}

Since direct preparation of magic states is noisy, magic state distillation is employed to convert many noisy magic states into fewer, higher-fidelity ones~\cite{bravyi2005universal}.
The 15-to-1 protocol based on Reed--Muller code is a widely used scheme, which consumes $15$ input and produces a single output.
Figure~\ref{fig:2:factory} shows a factory implementing this protocol.

When realized on the surface code, the output magic state error rate of a factory can be estimated~\cite{beverland2022assessing} by
\begin{equation}
\epsilon_{\text{out}} = 35 \,\epsilon_\text{in}^3 + 7.1\, p_L\,,
\label{eq:2:suppression}
\end{equation}
where $\epsilon_\text{in}$ is the input raw magic-state error rate and $p_L$ is the logical Clifford error rate in Equation~\eqref{eq:2:surface-error}.
A factory may also fail and discard its outputs, with success probability
\begin{equation}
p_{\text{succ}} = 1 - 15\,\epsilon_\text{in} - 356\, p_L\,.
\label{eq:2:success}
\end{equation}

\paragraph{Multi-level distillation}

Practical-scale quantum computers are expected to apply \emph{multi-level distillation}~\cite{jones2013multilevel} to achieve fidelities beyond what is attainable from only cubic suppression, as shown in Equation~\eqref{eq:2:suppression}.
For example, resource estimates~\cite{beverland2022assessing} show that quantum chemistry requires magic-state fidelities below $10^{-14}$, while superconducting devices typically generate raw magic states with error rates on the order of $10^{-4}$~\cite{beverland2022assessing}, necessitating two or three levels of distillation.
Table~\ref{tab:2:pipeline} illustrates an example of a three-level distillation pipeline, where each level employs a 15-to-1 factory.

However, although each round of distillation reduces the magic-state error rate cubically, this improvement is finally limited by the logical error rate $p_L$, which is not suppressed by distillation (Eq.~\eqref{eq:2:suppression}).
Therefore, the code distance $d$ must be increased across levels to reduce $p_L$ accordingly (Eq.~\eqref{eq:2:surface-error}) and to keep $\epsilon_\text{in}^3$ and $p_L$ within comparable regimes.

\begin{table}[htbp]
    \centering
    \caption{Example parameters for a three-level distillation pipeline.
    Code distance increases across levels, improving magic-state fidelity at the cost of higher qubit usage and longer execution time.
    Each level uses a 15-to-1 factory; lower levels must run multiple times to provide sufficient inputs to the higher levels.
    This table shows a typical input to our dynamic pipeline scheduling problem.
    }

    \label{tab:2:pipeline}
    \begin{tabular}{lccc}
        \toprule
        Distillation level & 1 & 2 & 3 \\
        \midrule
        Code distance          & 3                     & 9                     & 15 \\
        Input fidelity         & $1.0 \times 10^{-3}$  & $2.1 \times 10^{-3}$  & $2.5 \times 10^{-6}$ \\
        Output fidelity        & $2.1 \times 10^{-3}$  & $2.5 \times 10^{-6}$ & $2.1\times 10^{-9}$ \\
        Physical qubit         & 255 & 2415 & 6735 \\
        Execution time & $13.2~\mu\text{s}$ & $39.6~\mu\text{s}$ & $66.0~\mu\text{s}$ \\
        \bottomrule
    \end{tabular}
\end{table}

\section{Motivation and design insights}
\label{sec:motivation}

\subsection{Limitations of existing architectures}
\label{sub:3:existing}

\begin{figure}[t]
    \begin{center}
        \includegraphics[width=\columnwidth]{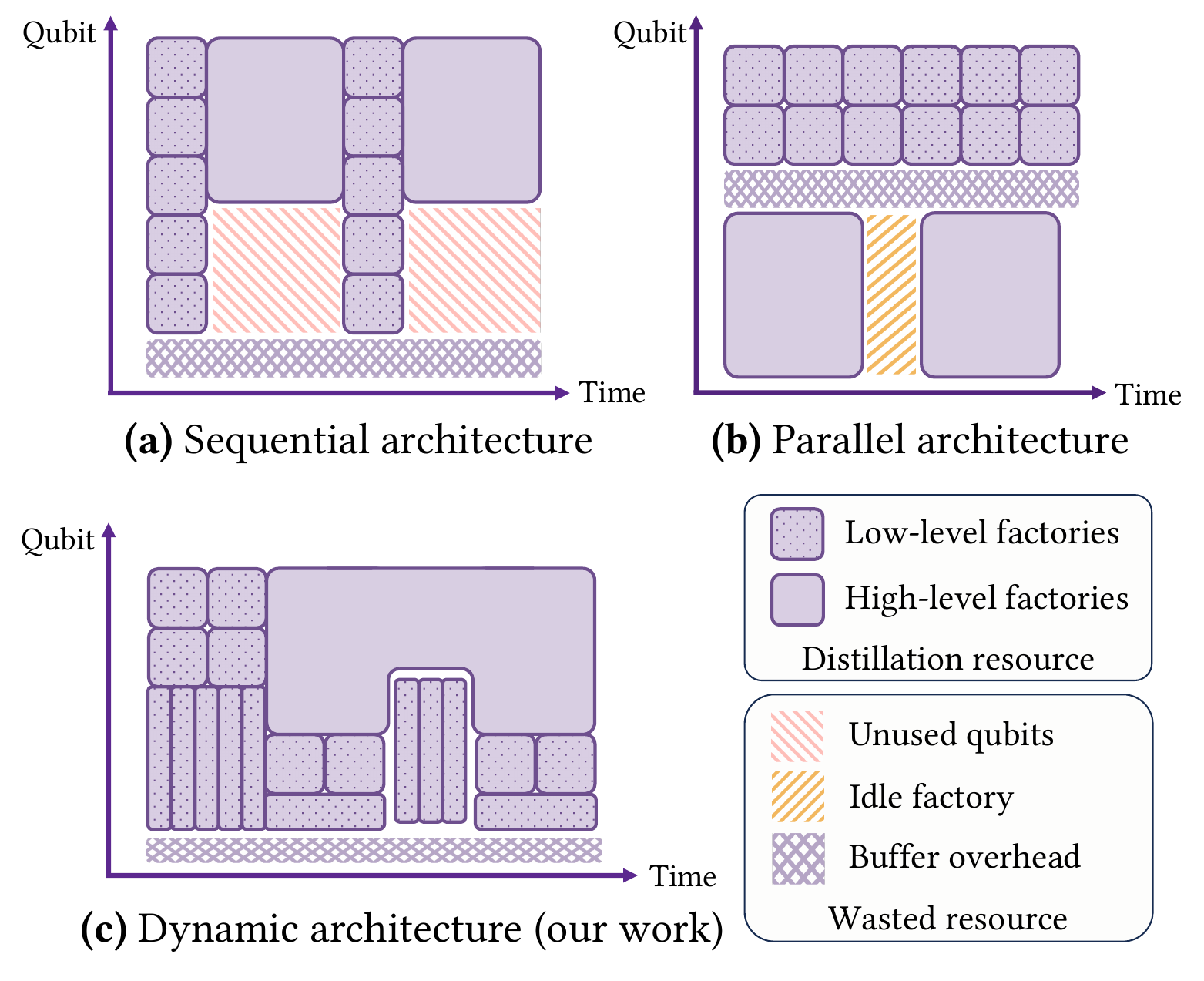}
        \caption{
        Resource underutilization for static methods and our work.
        Each square represents a factory execution, with width indicating execution time and height indicating qubit usage.
        \textbf{(a)} and \textbf{(b)} illustrate inefficiencies in unused qubits, idle factories, and buffer overhead in existing architectures~\protect{\cite{beverland2022assessing,silva2024optimizing}}.
        \textbf{(c)} Our work addresses these issues by enabling precise control over factory scheduling and resource allocation.
        }
        \label{fig:3:limitations}
    \end{center}
\end{figure}

In the sequential architecture~\cite{beverland2022assessing}, all physical qubits are allocated to the first level of distillation at the beginning.
Then, these qubits are reused for the second level and so on.
Although factory size increases with each level, the number of factories decreases.
As a result, higher-level factories typically either fail to fully utilize all available qubits (Fig.~\ref{fig:3:limitations}(a)) or occupy more qubits than lower-level factories.
Both situations lead to a substantial number of qubits unused, reducing the efficiency of magic-state production.

In the parallel architecture~\cite{silva2024optimizing}, all qubits are allocated to all factories which operate in parallel.
To balance production and consumption speed across levels, one must carefully tune the number of factories at each level.
However, due to the discrete execution time of factories, perfect matching is nearly impossible.
Excess magic-state production at a lower level factory leads to high buffer overhead, while insufficient production stalls higher-level factories and wastes their qubits (Fig.~\ref{fig:3:limitations}(b)).
These limitations reveal a key insight: \emph{fixing either the spatial or temporal structure of the pipeline hinders overall optimization opportunities.}

\subsection{Key insight: the burst-then-steady pattern}
\label{sec:3:litinski}

Litinski~\cite{litinski2019game} proposed a quantum circuit simplification technique that pushes all Clifford gates to the end of a circuit using gate commutation techniques, reducing the circuit to a sequence of non-Clifford rotations.
When applied to magic state distillation, every distillation protocol can be simplified to a sequence of rotations, each consuming one magic state~\cite{litinski2019magic}.
This technique reveals a key structural property of distillation circuits: each factory consumes magic states in a \emph{burst-then-steady} pattern, requiring multiple states initially followed by periodic single-state consumption.
For example, Figure~\ref{fig:2:factory} illustrates the simplified 15-to-1 distillation protocol, which consists of 11 non-Clifford rotations.
Four magic states are consumed for initialization, and 11 consecutive rotations are then steadily performed, each of which consumes one magic state.
This pattern changes the pipeline design paradigm from a \emph{prepare-then-execute} model to a \emph{supply-on-demand} model.
Instead of pre-preparing all required magic states, the pipeline can supply magic states on demand as factories consume them.

\subsection{Our approach: making the pipeline dynamic}

Based on the observation from the burst-then-steady consumption pattern, we propose a \emph{fully dynamic} pipeline architecture.
Our approach dynamically adjusts the pipeline structure based on available qubits and magic-state demand, allowing factories at different levels to execute at arbitrary times. We overcome the limitations of existing designs by carefully controlling the factory scheduling (Fig.~\ref{fig:3:limitations}(c)).
Resource waste from unused qubits and idle factories can be reduced by deploying additional factories using these residual and vacant qubits.
Buffer overhead can also be reduced by dynamically supplying magic states on demand, rather than preloading them.
Since the fully dynamic pipeline can adjust the factories in flexible ways, the existing sequential and parallel pipelines can be viewed as special cases of our more general framework.
This leads us to our core question: \emph{how do we construct an efficient dynamic pipeline?}
The challenge is to determine when each factory should execute and how many qubits should be allocated to each factory.

\section{Problem formulation}
\label{sec:problem}

\paragraph{Factory model}
We assume each factory has the following properties:
\begin{enumerate}[label=(\arabic*)]
    \item Input pattern: the number of required input magic states and the timesteps at which they are consumed in the distillation circuit. 
    \item Output: the number of output magic states and their fidelity as a function of input state fidelity.
    \item Resource cost: execution time and qubit usage.
    \item Success probability as a function of input state fidelity.
\end{enumerate}
This design allows us to ensure generality across distillation protocols and implementations, hiding low-level details such as lattice surgery operations. Techniques like physical layout optimization are beyond the scope of this work, but can be applied in a complementary fashion. 

\paragraph{Problem input}
We are given a sequence of distillation factories for each level, each of which has the model above. Following prior works on resource estimation~\cite{beverland2022assessing,harrigan2024qualtran}, we only consider a single type of factory at each level.
Thus, the input can be represented as a sequence of code distances, which determines other factory parameters.  Table \ref{tab:2:pipeline} shows an example of a $(3, 9, 15)$ pipeline with three rounds of distillation using 15-to-1 factories with code distances $3$, $9$ and $15$ at each level. In practice, application requirements (e.g. desired accuracy, magic-state count) are used to determine this input.
We note that buffer size is not assumed to be part of the input, as our architecture design selects the optimal buffer size automatically (see Sec.~\ref{sec:5:spatial}).

\paragraph{Optimization objectives and output.}
We construct the dynamic pipeline by scheduling factories and allocating qubit resources to factories. 
We aim to jointly optimise both space and time and produce a Pareto front ~\cite{ishizaka2013multi}, where each point represents a valid dynamic pipeline configuration. This subsumes work that focuses on either space (sequential pipelines) or time (parallel pipelines) minimization~\cite{silva2024optimizing, beverland2022assessing}.

\paragraph{Constraints.}
The distillation pipeline resembles a multi-level supply chain, with the fundamental constraint being that magic states must be produced before consumption by the next level.
Since perfect production--consumption synchronization introduces rigidity and failure vulnerability~\cite{hirano2024magicpool}, we employ buffers to temporarily store magic states between levels, which is a common practice in distillation architectures~\cite{hirano2024magicpool,silva2024optimizing}.
Through a combination of scheduling and buffer provisioning, we must ensure that buffer levels never become negative (demand outstrips production), maintaining feasible operation throughout the pipeline.

\section{Dynamic pipeline design}
\label{chap:design}

\subsection{Decomposition into two-level subproblems}
\label{sec:4:decomposition}

We begin by decomposing the multi-level factory scheduling problem into a sequence of subproblems, since solving it directly is challenging.
This problem resembles the NP-hard project scheduling problem under resource constraints~\cite{demeulemeester2002project}, and the multi-level nature of the pipeline adds additional complexity.
However, for multi-level pipelines, the magic states produced by the first $\ell$ levels can only be consumed by level-$(\ell{+}1)$ factories.
This locality allows us to perform scheduling independently between adjacent levels and compose the schedules. 

We start by constructing and optimizing the dynamic pipeline schedule for the first two levels. We obtain a Pareto front of pipeline configurations with their qubit and time usage, represented as qubit--time pairs $\{(Q_i, T_i)\}$.
To incorporate the third level, we treat the optimized first two levels as a single low-level factory and the third level as the high-level factory. 
For example, if both of the first two levels use 15-to-1 protocols, we view them as a single 225-to-1 factory while making decisions for the third level.
Since multiple two-level schedules exist with different qubit--time trade-offs, we can choose any combination of low-level factories to supply the magic states for the third-level factory.

In summary, the factory scheduling problem reduces to a recursive sequence of two-level subproblems, which can be solved from a supply-driven perspective, with low-level factories as producers and high-level factories as consumers:
\begin{itemize}
    \item \textbf{Input:} A set of low-level factories with qubit--time trade-offs $\{(Q_i, T_i)\}$ and one high-level factory.
    \item \textbf{Objective:} Select and schedule any combination of low-level factories to supply magic states for the high-level factory, minimising both qubit and time usage.
    \item \textbf{Output:} A new Pareto front $\{(Q'_i, T'_i)\}$, which corresponds to different schedules for the combined pipeline.
\end{itemize}

\subsection{Dynamic pipeline strategies}
\label{sec:5:overview}
To solve the two-level subproblem, we describe our strategies that enable dynamism in the pipeline, followed by the techniques for scheduling and qubit resource allocation. 

\paragraph{Key strategy 1: partial-input launch.}
To supply the initial burst demand, we first deploy as many low-level factories as possible to rapidly fill the buffer before launching the high-level factory.
After its deployment and launch, the remaining qubits are used to run additional low-level factories in parallel to maintain a steady supply.
Unlike traditional pipelines, this strategy allows the high-level factory to start before all inputs are prepared.

Can we avoid factory stalls solely through partial-input launch? A high-level factory stalls when the buffer becomes empty, which occurs when the production falls short of the consumption.
Although adjusting the launch time can reduce stall risk, complete elimination is rarely achievable.
Since distillation protocols use post-selection to filter erroneous magic states, distillation failures are unavoidable~\cite{litinski2019magic}.
Although enlarging the buffer can mitigate the impact of such failures, it incurs additional qubit overhead. Therefore, we require mechanisms to mitigate the impact of stalls.

\begin{figure}[tbp]
    \centering
    \includegraphics[width=\linewidth]{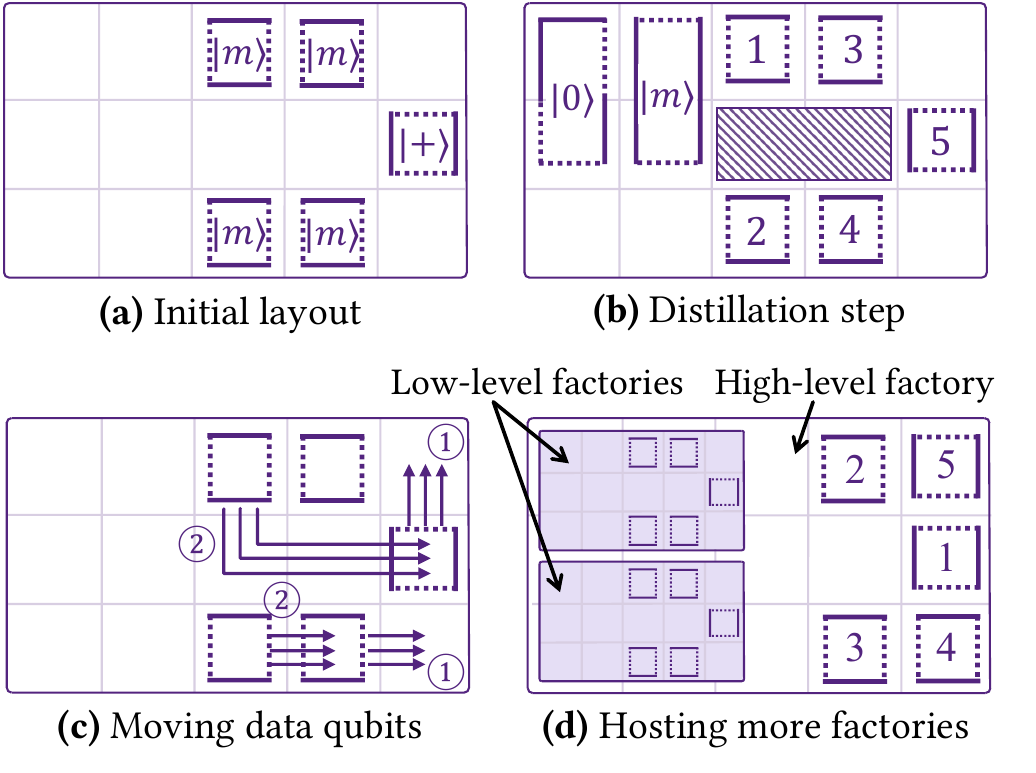}
    \caption{Ancilla reuse strategy using lattice surgery.
    \textbf{(a)} A typical layout of a 15-to-1 distillation factory~\protect{\cite{litinski2019game}} with only 5 logical data qubits (patches). Other patches can be safely reused during stalls without affecting the distillation process.
    \textbf{(b)} Each distillation step resets and uses ancilla qubits (shaded patches and the $|0\rangle$ patch) to implement a non-Clifford rotation, consuming one input magic state (the $|m\rangle$ patch).
    \textbf{(c)} When the high-level factory stalls, our ancilla reuse strategy moves data qubits aside to free up space, using two steps in this case. Circled numbers indicate move times.
    \textbf{(d)} The strategy then deploys two additional low-level factories by reusing the ancilla region of the high-level factory.
    }
    \label{fig:5:reuse}
\end{figure}

\paragraph{Key strategy 2: ancilla reuse during stalls.}
We can leverage idle qubits during stalls to run additional low-level factories to accelerate distillation.
In a typical implementation of the 15-to-1 distillation protocol (Fig.~\ref{fig:5:reuse}(a)), only 5 logical qubits store data, while the remaining 10 serve as temporary ancilla qubits.
These qubits are reset and reused in each non-Clifford rotation (Fig.~\ref{fig:5:reuse}(b)), and thus can be safely reused during stalls without disturbing the protocol.

To accommodate more low-level factories, we temporarily move the data qubits aside when a stall occurs (Fig.~\ref{fig:5:reuse}(c)).
These moves can be performed in parallel, and in lattice surgery each move costs one stabilizer measurement round, independent of distance~\cite{litinski2019game}.
Moving data qubits introduces minimal delay: most data qubits are moved in a single step, with a second step needed only if some paths overlap, and a third step is rarely necessary.
Once space is freed, we can deploy additional low-level factories in the vacated region to accelerate the magic-state production (Fig.~\ref{fig:5:reuse}(d)).

\paragraph{Three-phase execution.}
With these strategies, the overall execution of the two-level distillation pipeline proceeds in three phases.
\begin{enumerate*}[label=(\arabic*)]
    \item Distillation begins with all qubits dedicated to running low-level factories until the buffer accumulates enough magic states.
    \item The high-level factory is then launched, with the remaining qubits allocated to low-level factories running in parallel to provide a steady supply.
    \item When the high-level factory stalls due to insufficient input magic states, the system enters the third phase, reusing its ancilla qubits to run additional low-level factories.
    Once enough magic states are buffered, the high-level factory resumes, returning the system to the second phase.
\end{enumerate*}

There are two additional techniques required to complete our design.
Temporally, we design a factory scheduler that determines when to start factories and switch between execution phases.
Spatially, we implement a resource allocator that selects the optimal combination of low-level factories for each phase and determines the best buffer size.

\subsection{Factory scheduler}
\label{sec:5:temporal}
We present the design of a scheduler that keeps querying the buffer's magic-state count during execution and determines the launch and resumption time of the high-level factory.
The goal is to align the magic-state production with the consumption pattern, since the discrepancy between them leads to either excessive buffer requirements (production exceeds consumption), or frequent stalls (production falls short of consumption).
Specifically, we calculate buffer thresholds $N_\text{th}$ and $N_\text{th}'$ for launching and resuming the high-level factory, respectively.
Our scheduler triggers the corresponding action when the buffer count reaches these thresholds.

Two parameters that impact these decisions are the steady-state consumption rate of the high-level factory and the maximum production rate of the low-level factories.
The consumption rate is set by the distillation protocol. It equals the inverse of the time interval $\tau$ between consecutive magic-state consumptions, i.e. $R_\text{cons} = 1/\tau$.
Suppose the set of low-level factories is denoted by $\mathscr{F}$, the total production rate is then $R_\text{prod} = \sum_{f \in \mathscr{F}} M_f / T_f$, where $M_f$ is the number of magic states produced by factory $f$ and $T_f$ is the time taken to produce them.
If $R_\text{prod} \geq R_\text{cons}$, launching immediately after preparing the first burst inputs suffices.
If $R_\text{prod} < R_\text{cons}$, additional inputs must be pre-buffered to avoid stalls.

\begin{figure}[t]
    \centering
    \includegraphics[width=\columnwidth]{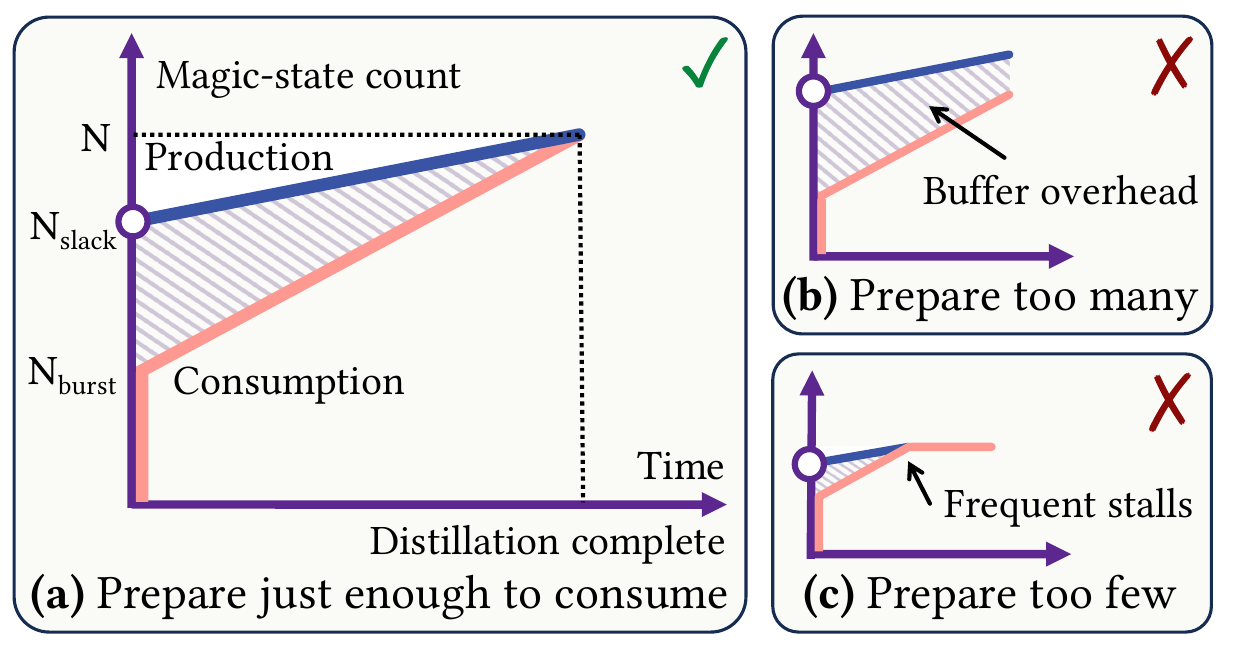}
    \caption{The method for determining the number of pre-buffered magic states.
    Two lines indicate the cumulative production and consumption count; the heights of shaded regions indicate the required buffer sizes.
    Consumption follows a burst-then-steady pattern, where $N$ is the total demand and $N_\text{burst}$ is the initial burst demand given by the protocol.
    We adjust the number of pre-buffered magic states $N_\text{slack}$ (circles on the y-axes) to keep production ahead of consumption while minimizing buffer overhead.
    \textbf{(a)} The optimal $N_\text{slack}$ is the minimum number of states required to compensate the production shortfall.
    \textbf{(b)} and \textbf{(c)} show suboptimal choices.
    }
    \label{fig:5:temporal}
\end{figure}

In the latter case, our strategy is to prepare slightly more magic states than the initial burst demand before launching the high-level factory, but not too many, as shown in Figure~\ref{fig:5:temporal}.
We pre-buffer just enough magic states to compensate for the production shortfall during execution (Fig.~\ref{fig:5:temporal}(a)).
Preparing more only increases buffer size without benefit (Fig.~\ref{fig:5:temporal}(b)), while preparing fewer leads to stalls (Fig.~\ref{fig:5:temporal}(c)).
Although we can rely on ancilla reuse to produce magic states during these stalls, its production rate is lower than normal execution, since data qubits for high-level factory cannot be reused, leading to fewer qubits available for low-level factories.

The required number of pre-buffered magic states is 
\begin{equation}
    N_\text{slack} = \max \left\{ N_\text{burst},\, N - \left\lfloor \frac{N - N_\text{burst}}{R_\text{cons}} R_\text{prod}\right\rfloor \right\}\,,
    \label{eq:5:temporal:N-slack}
\end{equation}
where $N$ is the total number of magic states required by the high-level factory and $N_\text{burst}$ is the initial burst demand.
The remaining $N - N_\text{burst}$ states are the steady-state demand.
Considering buffer capacity, the actual launch threshold is $N_\text{th} = \min \left\{ N_\text{slack}, N_\text{buf} \right\}$,
where $N_\text{buf}$ denotes the buffer size.

The resumption threshold is computed similarly, but the number of remaining magic-state demand depends on how many rotations are still pending in the high-level factory.
Let $n_\text{rot}$ denote this number, then the required number of pre-buffered magic states is
\begin{equation}
    N'_\text{slack} = \max \left\{ 1,\, n_{rot} - \left\lfloor \frac{n_{rot}}{R_\text{cons}} R_\text{prod}\right\rfloor \right\}\,.
    \label{eq:5:temporal:N-prime-slack}
\end{equation}
Considering the buffer size, the resumption threshold is  $N'_\text{th} = \min \left\{ N'_\text{slack},\, N_\text{buf} \right\}$.
Unlike the high-level factory launch threshold $N_\text{th}$ which could be computed statically, the resumption threshold $N'_\text{th}$ depends on the number of remaining rotations $n_\text{rot}$ and is computed dynamically at run time.

\subsection{Resource allocator}
\label{sec:5:spatial}
In this section, we discuss resource allocation in the two-level dynamic pipeline.
Our primary goal is to determine an appropriate combination of low-level factories to deploy for each execution phase and the buffer size.

In both the dedicated low-level factory phase (before the high-level factory launches) and the ancilla-reuse phase (after it stalls), the objective is to minimize the time to reach the buffer threshold (either the launch threshold $N_\text{th}$ or resumption threshold $N'_\text{th}$), subject to a fixed qubit budget.
This leads to a constrained bin-packing problem, which we formulate as an integer linear program.
Let $N_{\text{threshold}}$ denote the threshold and let $Q_{\text{max}}$ denote the available qubit budget for only low-level factories.
The problem can then be formulated as
\begin{align}
    \min &\qquad T \nonumber \\
    \text{subject to} &\qquad 
    \sum_{i=1}^{|\mathscr{F}|} n_i \cdot Q_i \leq Q_{\text{max}} \label{eq:5:spatial:ilp:budget} \\
    &\qquad \sum_{i=1}^{|\mathscr{F}|} n_i \cdot k_i \cdot M_i \geq N_{\text{threshold}} \label{eq:5:spatial:ilp:count} \\
    &\qquad k_i \cdot T_i \leq T \quad \forall i \in \{1, \cdots, |\mathscr{F}|\} \label{eq:5:spatial:time} \\
    &\qquad n_i, k_i \in \mathbb{Z}_{\geq 0}, \quad T \in \mathbb{Z}_{\geq 0} \label{eq:5:spatial:int}
\end{align}
where $Q_i, M_i, T_i$ are qubit usage, number of magic states produced, and time taken by the $i$-th factory, respectively.
The variables $n_i$ and $k_i$ represent the $i$-th factory is instantiated $n_i$ copies to run in parallel, each of which executes $k_i$ times within time $T$, as constrained by Equation~\eqref{eq:5:spatial:time}.
Equation~\eqref{eq:5:spatial:ilp:budget} ensures total qubit usage does not exceed the qubit budget, and Equation~\eqref{eq:5:spatial:ilp:count} ensures the target threshold is met.

When both low- and high-level factories are executing in parallel, there is no buffer threshold, as execution continues until the system stalls or completes.
The objective is instead to maximize the production rate under a fixed qubit budget $Q_\text{max}$.
Similarly, we formulate the optimization problem as
\begin{align}
    \max & \qquad R_\text{prod} \nonumber \\
    \text{subject to} & \qquad 
    \sum_{i=1}^{|\mathscr{F}|} n_i \cdot Q_i \leq Q_{\text{max}} \label{eq:5:spatial:prog:budget} \\
    & \qquad R_\text{prod} = \sum_{i=1}^{|\mathscr{F}|} n_i \cdot \frac{M_i}{T_i} \label{eq:5:spatial:prog:rate} \\
    & \qquad n_i \in \mathbb{Z}_{\geq 0}, \quad \forall i \in \{1, \ldots, |\mathscr{F}|\} 
\end{align}
where Equation~\eqref{eq:5:spatial:prog:budget} constraints the qubit budget, and Equation~\eqref{eq:5:spatial:prog:rate} calculates the production rate of magic states for the set of low-level factories.

These optimizations are fast and scalable. The first optimization can be solved using a standard ILP solver, with approximately 150 variables and 80 constraints for three-level pipelines, yielding a solution in 0.2 seconds. The second problem degenerates into a standard bin-packing problem, which can be solved efficiently via dynamic programming.

\paragraph{Optimize the buffer size}
Buffer size is another critical design parameter that directly impacts performance.
While buffering has been introduced into distillation pipelines in prior work~\cite{hirano2024magicpool,silva2024optimizing}, the choice of buffer size has not been considered.
We observe that buffering overhead is substantial, necessitating buffer-size optimization.
For example, a typical 15-to-1 factory shown in Figure~\ref{fig:5:reuse}(a) occupies 15 patches, and storing a single magic state occupies 1 patch.
Thus, buffering only $8$ magic states consumes more than half the space of the entire factory.
On the other hand, a smaller buffer leads to frequent stalls, impacting the overall time efficiency.

To trade off buffer overhead and time efficiency, we exhaustively search all candidate buffer sizes and evaluate performance via simulation.
The search space is small enough for enumeration, as buffer size must be at least the initial burst demand $N_\text{burst}$ of the high-level factory and at most its total demand $N$.
For the 15-to-1 protocol, this range is 4--15.

\subsection{Putting it all together}
\label{sec:5:together}

We integrate the factory scheduler and the resource allocator into a unified dynamic pipeline architecture.
The set of available low-level factories is computed recursively, incorporating one higher-level factory at each step.
The total qubit budget $Q$ and the buffer size $N_\text{buf}$ are tunable parameters used to explore space--time trade-offs.

Before execution, the scheduler statically determines the set of low-level factories to be deployed in the initial phase prior to launching the high-level factory, as well as those to run in parallel alongside the high-level factory after its launch.
The launch threshold $N_\text{th}$ for initiating the high-level factory is also determined at this stage.
The scheduler then instructs the quantum device to begin the distillation process with the selected low-level factories.
During execution, the scheduler monitors the buffer magic-state count.
Once it reaches the threshold $N_\text{th}$, it triggers the launch of the high-level factory, transitioning the system into the second phase where low- and high-level factories execute in parallel.
Each time the high-level factory stalls, the scheduler dynamically determines both the set of low-level factories to deploy while reusing ancilla qubits, and the resumption threshold $N'_\text{th}$ required to resume high-level distillation.
The system ends when the high-level factory completes.

\paragraph{Outlook for hardware deployment.}
Large-scale distillation pipelines target future quantum computers, as current hardware lacks the scale to support practical quantum applications or distillation~\cite{beverland2022assessing}.
We therefore discuss the prospective deployment method of our dynamic pipeline architecture on future systems.
Our proposed factory scheduler and resource allocator can be implemented on classical hardware, acting as a control unit that issues commands to the quantum computer and reacts to runtime conditions.
This forms a classical--quantum hybrid system, where the classical controller may be implemented using a CPU~\cite{ball2021software,zhu2019training}, FPGA~\cite{stefanazzi2022qick,qin2019fpga,xu2021qubic,hornibrook2015cryogenic}, or SoC~\cite{stanco2022versatile}.
Recent experiments already demonstrate the feasibility of classical control of quantum devices and indicate that low-latency communication between classical and quantum components is achievable~\cite{ding2024experimental,dadpour2025low,carrera2024combining}.

\section{Experimental setup}
\label{sec:setup}

\paragraph{Simulation}
We implemented a logical-cycle-accurate distillation pipeline simulator in Python, using \texttt{gurobi} version 12.0.3 as our ILP solver~\cite{gurobi}. The simulation proceeds in discrete time steps, with the stabilizer measurement cycle $T_\text{stab}$ as the time unit (see Sec.~\ref{sec:background}), so that we can model the behavior of factories with different code distances. 

Since magic-state factories can probabilistically fail, we require a technique to estimate their scheduling impact.
The low failure probability prevents us from directly simulating the failures; even under a pessimistic physical error rate $\epsilon = 10^{-3}$ and the smallest code distance $3$, the failure rate of a 15-to-1 factory remains below $0.2\%$~\cite{beverland2022assessing}.
Instead, we analytically estimate this delay and add it to the total execution time to ensure accuracy. 
These failures are modeled using a Markov chain, details are available in Appendix~\ref{sec:appendix}.

\paragraph{Physical parameters}
We mainly use superconducting qubit parameters~\cite{beverland2022assessing, jurcevic2021demonstration,arute2019supremacy} with a two-qubit gate time $T_\text{2q} = 50~\mu\mathrm{s}$ and a measurement time $T_\text{meas} = 100~\mu\mathrm{s}$. A single round of stabilizer measurement therefore takes $T_\text{stab}= 400~\mu\mathrm{s}$, which is the time unit for our simulation (Eq.~\eqref{eq:2:stab}).
This conversion allows us to translate simulation time into real-time units for interpretation. 
We set both the error rate of input raw magic states and physical gates to be $\epsilon = 10^{-4}$, which is a slightly optimistic estimation based on experimental implementations~\cite{arute2019supremacy,jurcevic2021demonstration,beverland2022assessing}. 
This setting affects the distillation levels and code distance choice to achieve the target fidelity.

\paragraph{Basic factory type}
Following the baselines, we use the 15-to-1 distillation protocol. We use its compact lattice surgery implementation introduced in Ref.~\cite{litinski2019game}.
Each factory occupies $15$ logical qubits, of which $5$ are data qubits. It requires an initial burst demand of $4$ magic states and a steady demand of $11$. Each non-Clifford rotation takes one logical step, so the total execution time without delay is $11$ logical steps. 

\paragraph{Distillation benchmarks}
We enumerate combinations of code distances from 3 to 47 to construct benchmarks for two- and three-level pipelines, which cover a wide range of application distillation scenarios.
Scalable applications on superconducting platforms~\cite{arute2019supremacy,kjaergaard2020superconducting,devoret2013superconducting} typically require two-level distillation, and other physical platforms, e.g. Majorana~\cite{karzig2017scalable}, may require three-level distillation~\cite{beverland2022assessing}.

To validate the practicality of our approach, we also evaluate on five large-scale applications, which capture the core areas of quantum computing~\cite{Preskill2018nisq,preskill2023quantum}.
These include quantum simulation~\cite{feynman2018simulating,buluta2009quantum}, represented by the Ising, Heisenberg, and Hubbard models~\cite{lloyd1996universal,low2017optimal,hatano2005finding}. We also include a quantum chemistry application~\cite{von2021quantum} and factoring~\cite{shor1999polynomial,gidney2021factor}.
All programs are taken from the Azure resource estimator~\cite{Microsoft_Azure_Quantum_Development,Azure_Quantum_Resource_Estimator}.

\paragraph{Baselines}
We compare our method with the sequential~\cite{beverland2022assessing} and parallel~\cite{silva2024optimizing} baselines.
Since the original works involve broader architectural concerns such as code distance selection and layout design, we reimplement the core distillation pipeline models to enable fair comparison.
For each architecture, we compute the number of physical qubits $Q$ and the time $T$ required to distill a single high-fidelity magic state.

For sequential architecture~\cite{beverland2022assessing}, the distillation levels run in sequence while reusing the same qubits.
Their work increases the number of low-level factories to $16$ to tolerate failures and ensure $>99\%$ success.
Using $n_\ell$ to demonstrate the number of factory copies of level $\ell$, we set $n_\ell = 16^{L-\ell}$.
The total qubit cost is the peak demand across all levels $Q_\text{seq} = \max_{\ell=1}^L \left\{n_\ell \cdot Q_\ell \right\}$, and the total time cost is the sum of execution times $T_\text{seq} = \sum_{\ell = 1}^L T_\ell, $
where $Q_\ell$ and $T_\ell$ are the physical qubit cost and runtime for a level-$\ell$ factory, respectively.
The routing time is ignored, as assumed by all approaches.

For parallel architecture~\cite{silva2024optimizing}, all levels run concurrently.
We fix the number of the highest-level factory $n_L = 1$ and, following their method, recursively compute the number of factory copies $n_\ell$ at each lower level by matching production and consumption rates as
$    n_{\ell - 1}  {M_{\ell - 1} {P}_{\ell - 1}}/{T_{\ell - 1}} = n_\ell  {N_{\ell}}/{T_\ell}$ for all $\ell = 2, \ldots, L,$
where $P_\ell, N_\ell, M_\ell,$ and $T_\ell$ denote the success probability, magic-state demand, output state count, and runtime of a level-$\ell$ factory, respectively.
The total qubit cost includes factory and buffer regions $Q_\text{par} = \sum_\ell n_\ell  \left(Q_\ell + B_\ell \right)$,
where $B_\ell$ is the physical qubit cost for buffer space at level $\ell$.
We follow their manually constructed buffer sizes and set $4$ logical qubits for the first level's buffer and $8$ logical qubits for other levels' buffers.
The total time cost is the runtime for one execution of the highest-level factory $T_\text{par} = T_L$.
This estimate relies on the optimistic assumption that the $L$-level pipeline operates continuously without stalls.
However, our analysis shows that stalls are inevitable in practice, making this estimate slightly overoptimistic.

\paragraph{Metrics}
Each baseline yields a specific $(Q, T)$ pair, while our approach produces a Pareto frontier of all feasible pipeline schedules.
To compare with baselines, we use the qubit--time or space--time volume $(Q \cdot T)$ as the main metric.

This metric has a clear physical interpretation.
Suppose a program requires $M_\text{prog}$ magic states and should complete within time $T_\text{prog}$, assuming no delay from magic state supply.
A distillation pipeline with cost $(Q, T)$ can produce $\lfloor T_\text{prog} / T \rfloor$ magic states within this time by continuous execution.
To meet the demand, at least
$
\left\lceil M_\text{prog} / \lfloor T_\text{prog} / T \rfloor \right\rceil
$
copies of pipelines must be deployed, resulting in a total qubit count
\begin{equation}
Q_\text{total} = Q \cdot \left\lceil \frac{M_\text{prog}}{\lfloor T_\text{prog} / T \rfloor } \right\rceil \approx (Q \cdot T) \cdot \frac{M_\text{prog}}{T_\text{prog}}\,,
\label{eq:6:total-qubit}
\end{equation}
which is approximately proportional to $(Q\cdot T)$.
Therefore, this metric fairly indicates the resource efficiency of a distillation pipeline.

\section{Results}
\subsection{Improvements on distillation benchmarks}

\paragraph{Two-level distillation pipeline.}
We first enumerate all code-distance pairs from 3 to 21 for two-level distillation pipelines and compare the results of our dynamic pipeline with the baselines.
Among the enumerated pairs, the best output error rate achieved is $2\times 10^{-23}$.

\begin{figure}[tbp]
   \centering
   \includegraphics[width=\columnwidth]{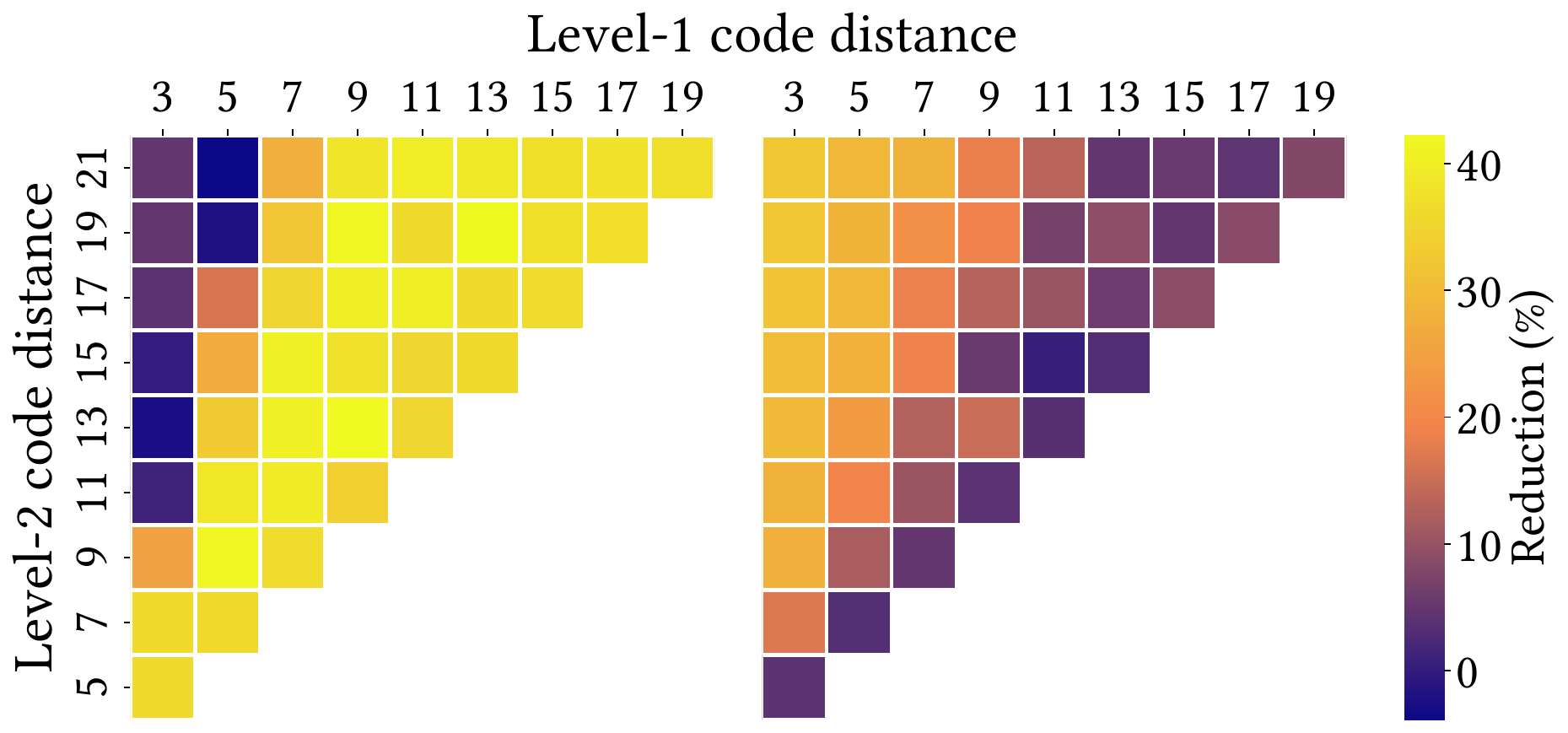}
    \captionsetup[sub]{skip=0pt}
    \begin{minipage}{0.4\columnwidth}
        \subcaption{Sequential~\protect{\cite{beverland2022assessing}}}
    \end{minipage}
    \begin{minipage}{0.4\columnwidth}
        \subcaption{Parallel~\protect{\cite{silva2024optimizing}}}
    \end{minipage}
    \begin{minipage}{0.2\columnwidth}
    \end{minipage}

    \caption{Qubit--time volume reductions for generated two-level pipeline benchmarks.
    Each cell corresponds to a benchmark, with code distances indicated on the axes; lighter colors denote larger reductions.}
   \label{fig:6:two-level}
\end{figure}

Figure~\ref{fig:6:two-level} shows the qubit--time volume reduction for each pair of code distances.
The results demonstrate our method reduces qubit--time volume across diverse code-distance combinations, achieving average improvements of 30\% and 15\% over the sequential and parallel baselines, respectively.
When code distances differ significantly (e.g., $(3,21)$), the high-level factory dominates both qubit and time cost, so improvements over the sequential baseline are limited.
When code distances are closer (e.g., $(5,7)$), the dynamic pipeline achieves substantial gains over the sequential baseline by avoiding unused qubits, while improvements over the parallel baseline are smaller due to reduced opportunities for ancilla reuse. 

Only 3 out of 90 benchmarks show a negative improvment, at worst up to $-4\%$, which we attribute to the baselines' overly optimistic evaluation assumptions.
The sequential baseline ignores its residual failure probability, while the parallel baseline assumes perfect synchronization without any production--consumption mismatches.
Our dynamic pipeline architecture subsumes both baselines as special cases, so in principle it should not underperform either of them.

Furthermore, code-distance combinations that yield such negative results are rarely used in practice.
We find that adjacent levels in these cases use distances that are either too close or too far apart.
However, a distillation level contributes to overall fidelity improvement only when the logical error rate (determined by the code distance) is reduced at the same pace as the cubic suppression of the magic-state error rate (see Sec.~\ref{sec:background}).
Consequently, only configurations with moderate code-distance growth across levels are practically effective and meaningful for evaluation.

\paragraph{Three-level distillation pipeline.}
To assess our architecture with more stringent fidelity requirements, we also consider three-level pipelines.
For charting results, we enumerate error-rate thresholds and compute the minimum required distillation qubit--time volumes.
Specifically, for each threshold, we search over all feasible code-distance sequences that achieve it and report the minimum qubit--time volume among the corresponding pipelines.
This approach also filters out impractical code-distance combinations, ensuring fair evaluation.

\begin{figure}[tbp]
    \centering
    \includegraphics[width=\columnwidth]{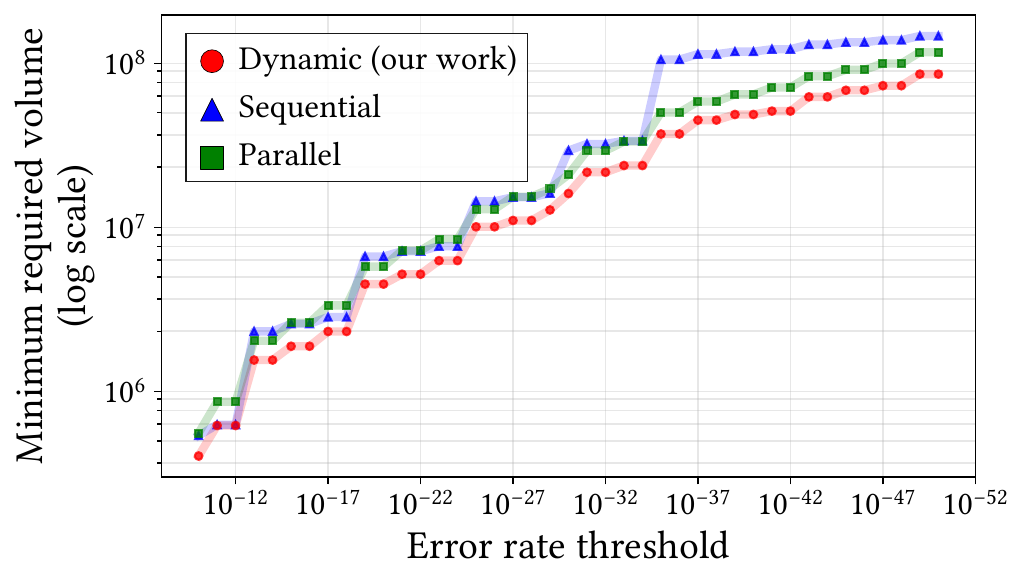}

    \caption{Minimum distillation qubit--time volumes required to meet various error-rate thresholds.
    For each threshold on the x-axis, we enumerate all feasible two- and three-level distillation pipelines that achieve it and report the minimum volume among them.
    Our method consistently reduces the required volume compared to both baselines~\protect{\cite{beverland2022assessing,silva2024optimizing}}, with the advantage becoming more pronounced under stricter error-rate requirements.
}
    \label{fig:6:three-level}
\end{figure}

Figure~\ref{fig:6:three-level} presents the results for error-rate thresholds ranging from $10^{-10}$ to $10^{-50}$.
It shows that our dynamic pipeline consistently achieves lower qubit--time volume across all thresholds.
The results are plotted on a logarithmic scale; in general, the stricter the error-rate threshold, the more pronounced the advantage of our method.
This trend highlights the scalability of our approach in meeting the stringent fidelity requirements of larger-scale applications.

Table~\ref{tab:6:stats} summarizes the reduction statistics across all evaluated error-rate thresholds.
Overall, our dynamic architecture achieves average reductions of 26\%--37\% in qubit--time volume.
For the sequential baseline, only two data points exhibit the worst-case reduction of 1\%, while all other cases show improvements of at least 18\%.
For the parallel baseline, our method achieves at least 22\% reduction.

\begin{table}[htbp]
    \small
    \caption{Statistics of qubit--time volume reductions over the baselines across error-rate thresholds from $10^{-10}$ to $10^{-50}$.}
    \label{tab:6:stats}
    \centering
    \begin{tabular}{lcccc}
        \toprule
        Baseline & Maximum & Minimum & Median & Average\\
        \midrule
        Sequential~\cite{beverland2022assessing} & 65\% &  1\%  &   33\%  &   37\%     \\
        Parallel~\cite{silva2024optimizing} &   31\%  & 22\%  &  26\%  & 26\%   \\
        \bottomrule
    \end{tabular}
\end{table}

\subsection{Improvements on applications}
We use Microsoft Azure resource estimator~\cite{Microsoft_Azure_Quantum_Development,Azure_Quantum_Resource_Estimator} to derive the distillation requirements for each application benchmark, i.e. magic-state demand, target fidelity, and execution time.
From these parameters, we determine the necessary distillation levels and code distances and then construct the dynamic pipeline.
We report results on both superconducting and Majorana devices using parameters given by Ref.~\cite{beverland2022assessing}.
While program execution could be slowed down to ease the demand on distillation factories and thereby reduce qubit cost~\cite{beverland2022assessing,silva2024optimizing,chatterjee2025q}, we assume no slowdown in order to isolate the impact of our pipeline design.
Table~\ref{tab:6:application-benchmarks} reports the results.

To compute the distillation qubit cost, we determine the number of pipeline copies required to meet the application's magic-state demand, assuming pipelines run continuously throughout the application execution.
Thus, both qubit and time savings from a single pipeline are already reflected in the reported qubit cost, as its shorter execution times reduce the number of required copies.

We observe significant percentage reductions in both distillation and total qubits (including program data qubits).
Focusing only on the distillation overhead, our method consistently reduces qubit usage, achieving 16\%--70\% reductions compared to the baselines.
When considering the total qubit usage, our approach still provides up to 70\% improvement for distillation-heavy applications, i.e. those with a large factory ratio.
Some applications (e.g. factoring) use only a very small percentage of qubits for distillation; we cannot obtain significant percentage reductions in total qubit usage in such cases.
However, even in these cases, the absolute reductions of $100\,\text{K}$--$2\,\text{M}$ qubits are offered by our method.

\begin{table*}[htbp]
\small
    \centering
    \caption{Distillation qubit cost comparison with sequential~\protect{\cite{beverland2022assessing}} and parallel~\protect{\cite{silva2024optimizing}} baselines on real-world applications. 
The reported distillation qubit cost reflects both qubit and time reductions of a single pipeline, as its shorter execution time reduces the number of required copies.
Code distances are selected to meet the required fidelity, and the factory ratio is the percentage of qubits used for distillation under our architecture.
Results for both superconducting and Majorana platforms are shown, using parameters from Ref.~\protect{\cite{beverland2022assessing}}.
Our method yields substantial qubit savings, especially for distillation-heavy applications.
}

    \label{tab:6:application-benchmarks}
    \begin{tabular}{cp{2.2cm}cccrrrcc}
        \toprule
        \multicolumn{2}{c}{\multirow[b]{2}{*}{\makecell{Application and \\ quantum platform}}} & \multicolumn{3}{c}{Application requirements} & \multicolumn{3}{c}{Distillation qubit cost} & \multicolumn{2}{c}{\makecell{\textbf{Distillation (total)} \\ \textbf{qubit reduction}}}\\
        \cmidrule(lr){3-5} \cmidrule(lr){6-8} \cmidrule(lr){9-10}
        & & \makecell{Required \\fidelity} & \makecell{Factory \\ratio} & \makecell{Code \\distances} & \makecell{Sequential} & Parallel & Dynamic & \textbf{Sequential} & \textbf{Parallel} \\
        \midrule
        \multirow{5}{*}{\rotatebox{90}{Supercond.}} & 
        Ising $10\times10$ & $7\times10^{-10}$ & 99.6\% & (3, 9) & 6,989,894 & 7,248,161 & 5,254,082 & \textbf{25\% (25\%)} & \textbf{28\% (27\%)} \\
        & Heis $40\times40$ & $2\times10^{-16}$ & 90\% & (5, 15) & 66,488,328 & 67,476,324 & 48,537,673 & \textbf{27\% (25\%)} & \textbf{28\% (26\%)} \\
        & Hub $40\times40$ & $2\times10^{-17}$ & 71\% & (5, 17) & 74,401,448 & 87,636,232 & 62,143,840 & \textbf{16\% (12\%)} & \textbf{29\% (23\%)} \\
        & Chemistry & $5\times10^{-18}$ & 66\% & (5, 17) & 2,888,587 & 3,402,419 & 2,412,694 & \textbf{16\% (12\%)} & \textbf{29\% (21\%)} \\
        & Factoring 2048 & $2\times10^{-17}$ & 4\% & (5, 17) & 2,489,885 & 2,932,794 & 2,079,677 & \textbf{16\% (1\%)} & \textbf{29\% (2\%)} \\
        \midrule
        \multirow{5}{*}{\rotatebox{90}{Majorana}} & 
        Ising $10\times10$ & $7\times10^{-10}$ & 99.6\% & (1, 5, 13) & 30,678,906 & 29,235,345 & 20,512,298 & \textbf{33\% (33\%)} & \textbf{30\% (30\%)} \\
        & Heis $40\times40$ & $2\times10^{-16}$ & 99.3\% & (5, 9, 23) & 1,968,054,489 & 707,952,275 & 591,710,478 & \textbf{70\% (70\%)} & \textbf{16\% (16\%)} \\
        & Hub $40\times40$ & $2\times10^{-17}$ & 97\% & (5, 9, 23) & 2,002,075,303 & 720,190,306 & 601,939,093 & \textbf{70\% (69\%)} & \textbf{16\% (16\%)} \\
        & Chemistry & $5\times10^{-18}$ & 40\% & (1, 3, 9) & 559,058 & 622,299 & 445,404 & \textbf{20\% (9\%)} & \textbf{28\% (14\%)} \\
        & Factoring 2048 & $2\times10^{-17}$ & 12\% & (1, 7, 21) & 6,334,318 & 6,813,431 & 4,729,303 & \textbf{25\% (4\%)} & \textbf{31\% (5\%)} \\
        \bottomrule
    \end{tabular}
\end{table*}

\begin{figure}[tb]
   \centering
   \includegraphics[width=0.97\columnwidth]{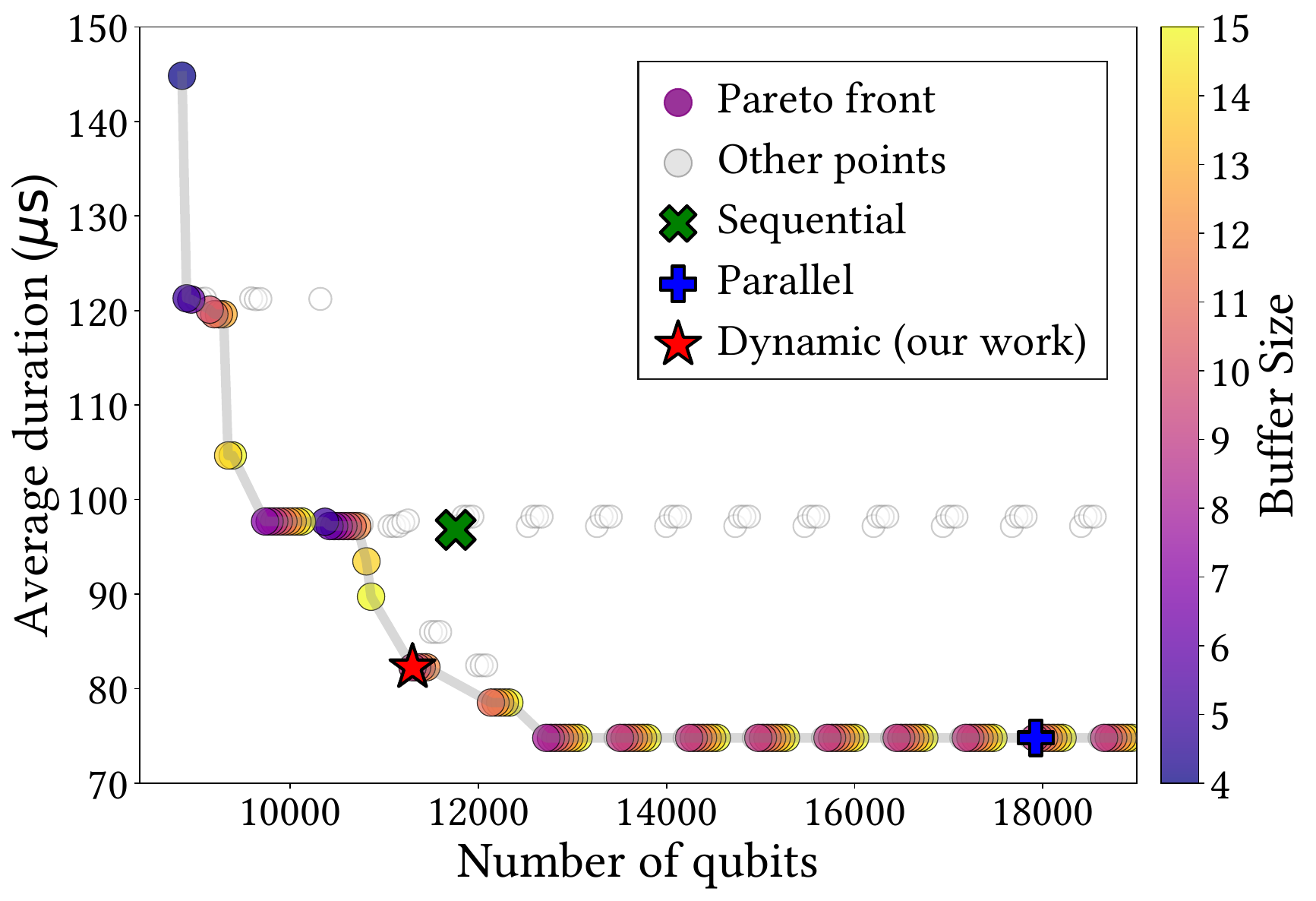}
   \caption{Space--time trade-off curve for a factory distilling a single magic state in the Hubbard benchmark.
    Each point represents a valid pipeline configuration under our architecture, with colored points showing the Pareto front (lighter colors denote larger buffer sizes). 
    While each baseline corresponds to a single fixed point, our method exposes the full trade-off space. 
    It also achieves the best qubit--time volume (star mark) with an optimal buffer size choice (10 in this case), outperforming both baselines.
   }
   \label{fig:6:chemistry}
\end{figure}

\paragraph{Case study of the Hubbard application.}
Here we investigate how our architecture enables full space--time trade-off opportunities in distillation pipelines with code distances $(5, 17)$ as used in the Hubbard model.
Figure~\ref{fig:6:chemistry} shows the Pareto front generated by our method to distill one high-fidelity magic state, with each point representing a feasible pipeline configuration.
The red star indicates the minimum space--time volume across all pipeline configurations, which occurs at a buffer size of $10$.

As expected, a smaller qubit budget results in longer execution times.
With limited qubits, only one high-level factory and a minimal buffer of size 4 (darkest color) can be deployed, forcing heavy ancilla reuse and leading to long runtimes.
As the budget increases, larger buffers (lighter colors) and additional factories can run in parallel, reducing execution time.
Eventually, the time cost plateaus at around $27\,\mu\mathrm{s}$, which equals the execution time of a single second-level factory, indicating that this factory has become the bottleneck.
It marks the optimal execution time, matching the parallel baseline.

Compared to the sequential baseline, we reduce both qubits and execution time.
Compared to the parallel baseline, which attains the minimal time usage, our method requires fewer qubits.
Our architecture also exposes the full trade-off space, creating opportunities for further compiler-level optimizations in FTQC schemes.
Moreover, the resource allocator explicitly considers buffer size and selects $10$ as the optimal value, providing practical guidance for buffer size selection.

\section{Related work}

\paragraph{Protocols for realizing non-Clifford gates}
Universal fault-tolerant quantum computing depends on the efficient implementation of non-Clifford gates~\cite{gottesman1998heisenberg}.
In Clifford+T (magic state) framework, fundamental protocols for T state distillation have been extensively studied~\cite{bravyi2012magic,meier2012magic,campbell2012magic,campbell2014enhanced,haah2018codes,litinski2019magic,chamberland2022building,knill2004fault, haah2017magic,lee2025low,hastings2018distillation,haah2017magicintermediate}, with further improvements on raw input states via zero-level distillation~\cite{hirano2024leveraging,itogawa2024even,singh2022high} and injection~\cite{li2015magic,lao2022magic,gidney2023cleaner}.
Other methods aim to bypass distillation, including catalysis~\cite{gidney2019efficient}, cultivation~\cite{gidney2024cultivation,vaknin2025magic}, and transversal-CNOT-based protocols~\cite{wan2024iterative,wan2024constant}.
Beyond the Clifford+T framework, there are alternative routes to universality such as Clifford+Rz (arbitrary rotation)~\cite{akahoshi2024partially,choi2023fault,sethi2025rescq} and code switching~\cite{beverland2021cost,heussen2024efficient,pogorelov2025experimental,butt2024fault}.
These approaches either impose stricter requirements on hardware or underlying error correction codes, or support only limited-fidelity non-Clifford operations, and thus do not eliminate the need for distillation.

\paragraph{Compilation of distillation factories}
Considerable effort has been devoted to compiling these distillation protocols onto specific quantum error correction schemes, especially based on the surface code.
One line of work focuses on the realization of a single factory, employing methods that range from manual optimization~\cite{fowler2018low,litinski2019game,litinski2019magic,prabhu2022new,fowler2013surface} to SAT-based automated approaches~\cite{tan2024sat}.
Another line of work addresses multiple factories, including factories placement~\cite{holmes2019resource,hirano2025locality}, inter-factory routing~\cite{ding2018magic} and buffering~\cite{hirano2024magicpool}.
These works are largely complementary to ours, as our architecture is compatible with all lattice-surgery-based distillation factory implementations, while delegating physical layout and routing tasks to the compiler.

\paragraph{Operating system supports for FTQC}
Our work resembles operating system support for resource management and task scheduling in a quantum computing environment.
Quantum system works have explored multi-program scheduling~\cite{giortamis2024orchestrating} and dynamic resource allocation~\cite{kan2025sparo}, but these approaches treat distillation factories as black-box components.
In contrast, we target the distillation process itself and offer flexible control over distillation factories.

\paragraph{Quantum circuit optimizers}
Methods for optimizing quantum circuits to reduce resource overhead have been widely explored via rewrite rules~\cite{hietala2021verified,kissinger2019pyzx,xu2023synthesizing}, unitary transformations~\cite{amy2013meet, patel2022quest}, their combinations~\cite{xu2025optimizing}, and qubit reuse strategies ~\cite{hua2023caqr,paler2016wire,jiang2024qubit}.
While such methods can also be applied to distillation circuits, our work operates at a higher level of abstraction, focusing on the dynamic scheduling of multi-level distillation rather than circuit-level optimization.

\section{Conclusions}
\label{sec:conclusion}

We have presented a dynamic pipeline architecture for multi-level distillation that allows the pipeline structure to evolve over time.
Through dynamic scheduling and resource allocation, our approach orchestrates the executions of factories across levels to improve resource utilization.
Compared with state-of-the-art architectures, our approach achieves significant reductions in distillation overhead and unlocks further optimization opportunities by offering flexible space--time trade-offs.
These advancements bring us closer to efficient and scalable fault-tolerant quantum computing systems.

\bibliographystyle{ACM-Reference-Format}
\bibliography{main}


\begin{thebibliography}{92}


\ifx \showCODEN    \undefined \def \showCODEN     #1{\unskip}     \fi
\ifx \showISBNx    \undefined \def \showISBNx     #1{\unskip}     \fi
\ifx \showISBNxiii \undefined \def \showISBNxiii  #1{\unskip}     \fi
\ifx \showISSN     \undefined \def \showISSN      #1{\unskip}     \fi
\ifx \showLCCN     \undefined \def \showLCCN      #1{\unskip}     \fi
\ifx \shownote     \undefined \def \shownote      #1{#1}          \fi
\ifx \showarticletitle \undefined \def \showarticletitle #1{#1}   \fi
\ifx \showURL      \undefined \def \showURL       {\relax}        \fi
\providecommand\bibfield[2]{#2}
\providecommand\bibinfo[2]{#2}
\providecommand\natexlab[1]{#1}
\providecommand\showeprint[2][]{arXiv:#2}

\bibitem[Acharya et~al\mbox{.}(2023)]%
        {google2023suppressing}
\bibfield{author}{\bibinfo{person}{Rajeev Acharya}, \bibinfo{person}{Igor
  Aleiner}, \bibinfo{person}{Richard Allen}, \bibinfo{person}{Trond~I
  Andersen}, \bibinfo{person}{Markus Ansmann}, \bibinfo{person}{Frank Arute},
  \bibinfo{person}{Kunal Arya}, \bibinfo{person}{Abraham Asfaw},
  \bibinfo{person}{Juan Atalaya}, \bibinfo{person}{Ryan Babbush},
  \bibinfo{person}{Dave Bacon}, \bibinfo{person}{Joseph~C Bardin},
  \bibinfo{person}{Joao Basso}, \bibinfo{person}{Andreas Bengtsson},
  \bibinfo{person}{Sergio Boixo}, \bibinfo{person}{Gina Bortoli},
  \bibinfo{person}{Alexandre Bourassa}, \bibinfo{person}{Jenna Bovaird},
  \bibinfo{person}{Leon Brill}, \bibinfo{person}{Michael Broughton},
  \bibinfo{person}{Bob~B Buckley}, \bibinfo{person}{David~A Buell},
  \bibinfo{person}{Tim Burger}, \bibinfo{person}{Brian Burkett},
  \bibinfo{person}{Nicholas Bushnell}, \bibinfo{person}{Yu Chen},
  \bibinfo{person}{Zijun Chen}, \bibinfo{person}{Ben Chiaro},
  \bibinfo{person}{Josh Cogan}, \bibinfo{person}{Roberto Collins},
  \bibinfo{person}{Paul Conner}, \bibinfo{person}{William Courtney},
  \bibinfo{person}{Alexander~L Crook}, \bibinfo{person}{Ben Curtin},
  \bibinfo{person}{Dripto~M Debroy}, \bibinfo{person}{Alexander Del
  Toro~Barba}, \bibinfo{person}{Sean Demura}, \bibinfo{person}{Andrew
  Dunsworth}, \bibinfo{person}{Daniel Eppens}, \bibinfo{person}{Catherine
  Erickson}, \bibinfo{person}{Lara Faoro}, \bibinfo{person}{Edward Farhi},
  \bibinfo{person}{Reza Fatemi}, \bibinfo{person}{Leslie Flores~Burgos},
  \bibinfo{person}{Ebrahim Forati}, \bibinfo{person}{Austin~G Fowler},
  \bibinfo{person}{Brooks Foxen}, \bibinfo{person}{William Giang},
  \bibinfo{person}{Craig Gidney}, \bibinfo{person}{Dar Gilboa},
  \bibinfo{person}{Marissa Giustina}, \bibinfo{person}{Alejandro Grajales~Dau},
  \bibinfo{person}{Jonathan~A Gross}, \bibinfo{person}{Steve Habegger},
  \bibinfo{person}{Michael~C Hamilton}, \bibinfo{person}{Matthew~P Harrigan},
  \bibinfo{person}{Sean~D Harrington}, \bibinfo{person}{Oscar Higgott},
  \bibinfo{person}{Jeremy Hilton}, \bibinfo{person}{Markus Hoffmann},
  \bibinfo{person}{Sabrina Hong}, \bibinfo{person}{Trent Huang},
  \bibinfo{person}{Ashley Huff}, \bibinfo{person}{William~J Huggins},
  \bibinfo{person}{Lev~B Ioffe}, \bibinfo{person}{Sergei~V Isakov},
  \bibinfo{person}{Justin Iveland}, \bibinfo{person}{Evan Jeffrey},
  \bibinfo{person}{Zhang Jiang}, \bibinfo{person}{Cody Jones},
  \bibinfo{person}{Pavol Juhas}, \bibinfo{person}{Dvir Kafri},
  \bibinfo{person}{Kostyantyn Kechedzhi}, \bibinfo{person}{Julian Kelly},
  \bibinfo{person}{Tanuj Khattar}, \bibinfo{person}{Mostafa Khezri},
  \bibinfo{person}{M{\'a}ria Kieferov{\'a}}, \bibinfo{person}{Seon Kim},
  \bibinfo{person}{Alexei Kitaev}, \bibinfo{person}{Paul~V Klimov},
  \bibinfo{person}{Andrey~R Klots}, \bibinfo{person}{Alexander~N Korotkov},
  \bibinfo{person}{Fedor Kostritsa}, \bibinfo{person}{John~Mark Kreikebaum},
  \bibinfo{person}{David Landhuis}, \bibinfo{person}{Pavel Laptev},
  \bibinfo{person}{Kim-Ming Lau}, \bibinfo{person}{Lily Laws},
  \bibinfo{person}{Joonho Lee}, \bibinfo{person}{Kenny Lee},
  \bibinfo{person}{Brian~J Lester}, \bibinfo{person}{Alexander Lill},
  \bibinfo{person}{Wayne Liu}, \bibinfo{person}{Aditya Locharla},
  \bibinfo{person}{Erik Lucero}, \bibinfo{person}{Fionn~D Malone},
  \bibinfo{person}{Jeffrey Marshall}, \bibinfo{person}{Orion Martin},
  \bibinfo{person}{Jarrod~R McClean}, \bibinfo{person}{Trevor McCourt},
  \bibinfo{person}{Matt McEwen}, \bibinfo{person}{Anthony Megrant},
  \bibinfo{person}{Bernardo Meurer~Costa}, \bibinfo{person}{Xiao Mi},
  \bibinfo{person}{Kevin~C Miao}, \bibinfo{person}{Masoud Mohseni},
  \bibinfo{person}{Shirin Montazeri}, \bibinfo{person}{Alexis Morvan},
  \bibinfo{person}{Emily Mount}, \bibinfo{person}{Wojciech Mruczkiewicz},
  \bibinfo{person}{Ofer Naaman}, \bibinfo{person}{Matthew Neeley},
  \bibinfo{person}{Charles Neill}, \bibinfo{person}{Ani Nersisyan},
  \bibinfo{person}{Hartmut Neven}, \bibinfo{person}{Michael Newman},
  \bibinfo{person}{Jiun~How Ng}, \bibinfo{person}{Anthony Nguyen},
  \bibinfo{person}{Murray Nguyen}, \bibinfo{person}{Murphy~Yuezhen Niu},
  \bibinfo{person}{Thomas~E O'Brien}, \bibinfo{person}{Alex Opremcak},
  \bibinfo{person}{John Platt}, \bibinfo{person}{Andre Petukhov},
  \bibinfo{person}{Rebecca Potter}, \bibinfo{person}{Leonid~P Pryadko},
  \bibinfo{person}{Chris Quintana}, \bibinfo{person}{Pedram Roushan},
  \bibinfo{person}{Nicholas~C Rubin}, \bibinfo{person}{Negar Saei},
  \bibinfo{person}{Daniel Sank}, \bibinfo{person}{Kannan Sankaragomathi},
  \bibinfo{person}{Kevin~J Satzinger}, \bibinfo{person}{Henry~F Schurkus},
  \bibinfo{person}{Christopher Schuster}, \bibinfo{person}{Michael~J Shearn},
  \bibinfo{person}{Aaron Shorter}, \bibinfo{person}{Vladimir Shvarts},
  \bibinfo{person}{Jindra Skruzny}, \bibinfo{person}{Vadim Smelyanskiy},
  \bibinfo{person}{W~Clarke Smith}, \bibinfo{person}{George Sterling},
  \bibinfo{person}{Doug Strain}, \bibinfo{person}{Marco Szalay},
  \bibinfo{person}{Alfredo Torres}, \bibinfo{person}{Guifre Vidal},
  \bibinfo{person}{Benjamin Villalonga}, \bibinfo{person}{Catherine
  Vollgraff~Heidweiller}, \bibinfo{person}{Theodore White},
  \bibinfo{person}{Cheng Xing}, \bibinfo{person}{Z~Jamie Yao},
  \bibinfo{person}{Ping Yeh}, \bibinfo{person}{Juhwan Yoo},
  \bibinfo{person}{Grayson Young}, \bibinfo{person}{Adam Zalcman},
  \bibinfo{person}{Yaxing Zhang}, \bibinfo{person}{Ningfeng Zhu}, {and}
  \bibinfo{person}{{Google Quantum AI}}.} \bibinfo{year}{2023}\natexlab{}.
\newblock \showarticletitle{Suppressing quantum errors by scaling a surface
  code logical qubit}.
\newblock \bibinfo{journal}{\emph{Nature}} \bibinfo{volume}{614},
  \bibinfo{number}{7949} (\bibinfo{date}{Feb.} \bibinfo{year}{2023}),
  \bibinfo{pages}{676--681}.
\newblock
\urldef\tempurl%
\url{https://doi.org/10.1038/s41586-022-05434-1}
\showURL{%
\tempurl}


\bibitem[Akahoshi et~al\mbox{.}(2024)]%
        {akahoshi2024partially}
\bibfield{author}{\bibinfo{person}{Yutaro Akahoshi}, \bibinfo{person}{Kazunori
  Maruyama}, \bibinfo{person}{Hirotaka Oshima}, \bibinfo{person}{Shintaro
  Sato}, {and} \bibinfo{person}{Keisuke Fujii}.}
  \bibinfo{year}{2024}\natexlab{}.
\newblock \showarticletitle{Partially fault-tolerant quantum computing
  architecture with error-corrected clifford gates and space-time efficient
  analog rotations}.
\newblock \bibinfo{journal}{\emph{PRX quantum}} \bibinfo{volume}{5},
  \bibinfo{number}{1} (\bibinfo{year}{2024}), \bibinfo{pages}{010337}.
\newblock
\urldef\tempurl%
\url{https://doi.org/10.1103/PRXQuantum.5.010337}
\showURL{%
\tempurl}


\bibitem[Amy et~al\mbox{.}(2013)]%
        {amy2013meet}
\bibfield{author}{\bibinfo{person}{Matthew Amy}, \bibinfo{person}{Dmitri
  Maslov}, \bibinfo{person}{Michele Mosca}, {and} \bibinfo{person}{Martin
  Roetteler}.} \bibinfo{year}{2013}\natexlab{}.
\newblock \showarticletitle{A meet-in-the-middle algorithm for fast synthesis
  of depth-optimal quantum circuits}.
\newblock \bibinfo{journal}{\emph{IEEE Transactions on Computer-Aided Design of
  Integrated Circuits and Systems}} \bibinfo{volume}{32}, \bibinfo{number}{6}
  (\bibinfo{year}{2013}), \bibinfo{pages}{818--830}.
\newblock
\urldef\tempurl%
\url{https://doi.org/10.1109/TCAD.2013.2244643}
\showURL{%
\tempurl}


\bibitem[Arute et~al\mbox{.}(2019)]%
        {arute2019supremacy}
\bibfield{author}{\bibinfo{person}{Frank Arute}, \bibinfo{person}{Kunal Arya},
  \bibinfo{person}{Ryan Babbush}, \bibinfo{person}{Dave Bacon},
  \bibinfo{person}{Joseph~C. Bardin}, \bibinfo{person}{Rami Barends},
  \bibinfo{person}{Rupak Biswas}, \bibinfo{person}{Sergio Boixo},
  \bibinfo{person}{Fernando G. S.~L. Brandao}, \bibinfo{person}{David~A.
  Buell}, \bibinfo{person}{Brian Burkett}, \bibinfo{person}{Yu Chen},
  \bibinfo{person}{Zijun Chen}, \bibinfo{person}{Ben Chiaro},
  \bibinfo{person}{Roberto Collins}, \bibinfo{person}{William Courtney},
  \bibinfo{person}{Andrew Dunsworth}, \bibinfo{person}{Edward Farhi},
  \bibinfo{person}{Brooks Foxen}, \bibinfo{person}{Austin Fowler},
  \bibinfo{person}{Craig Gidney}, \bibinfo{person}{Marissa Giustina},
  \bibinfo{person}{Rob Graff}, \bibinfo{person}{Keith Guerin},
  \bibinfo{person}{Steve Habegger}, \bibinfo{person}{Matthew~P. Harrigan},
  \bibinfo{person}{Michael~J. Hartmann}, \bibinfo{person}{Alan Ho},
  \bibinfo{person}{Markus Hoffmann}, \bibinfo{person}{Trent Huang},
  \bibinfo{person}{Travis~S. Humble}, \bibinfo{person}{Sergei~V. Isakov},
  \bibinfo{person}{Evan Jeffrey}, \bibinfo{person}{Zhang Jiang},
  \bibinfo{person}{Dvir Kafri}, \bibinfo{person}{Kostyantyn Kechedzhi},
  \bibinfo{person}{Julian Kelly}, \bibinfo{person}{Paul~V. Klimov},
  \bibinfo{person}{Sergey Knysh}, \bibinfo{person}{Alexander Korotkov},
  \bibinfo{person}{Fedor Kostritsa}, \bibinfo{person}{David Landhuis},
  \bibinfo{person}{Mike Lindmark}, \bibinfo{person}{Erik Lucero},
  \bibinfo{person}{Dmitry Lyakh}, \bibinfo{person}{Salvatore Mandr{\`a}},
  \bibinfo{person}{Jarrod~R. McClean}, \bibinfo{person}{Matthew McEwen},
  \bibinfo{person}{Anthony Megrant}, \bibinfo{person}{Xiao Mi},
  \bibinfo{person}{Kristel Michielsen}, \bibinfo{person}{Masoud Mohseni},
  \bibinfo{person}{Josh Mutus}, \bibinfo{person}{Ofer Naaman},
  \bibinfo{person}{Matthew Neeley}, \bibinfo{person}{Charles Neill},
  \bibinfo{person}{Murphy~Yuezhen Niu}, \bibinfo{person}{Eric Ostby},
  \bibinfo{person}{Andre Petukhov}, \bibinfo{person}{John~C. Platt},
  \bibinfo{person}{Chris Quintana}, \bibinfo{person}{Eleanor~G. Rieffel},
  \bibinfo{person}{Pedram Roushan}, \bibinfo{person}{Nicholas~C. Rubin},
  \bibinfo{person}{Daniel Sank}, \bibinfo{person}{Kevin~J. Satzinger},
  \bibinfo{person}{Vadim Smelyanskiy}, \bibinfo{person}{Kevin~J. Sung},
  \bibinfo{person}{Matthew~D. Trevithick}, \bibinfo{person}{Amit Vainsencher},
  \bibinfo{person}{Benjamin Villalonga}, \bibinfo{person}{Theodore White},
  \bibinfo{person}{Z.~Jamie Yao}, \bibinfo{person}{Ping Yeh},
  \bibinfo{person}{Adam Zalcman}, \bibinfo{person}{Hartmut Neven}, {and}
  \bibinfo{person}{John~M. Martinis}.} \bibinfo{year}{2019}\natexlab{}.
\newblock \showarticletitle{Quantum supremacy using a programmable
  superconducting processor}.
\newblock \bibinfo{journal}{\emph{Nature}} \bibinfo{volume}{574},
  \bibinfo{number}{7779} (\bibinfo{date}{01 Oct} \bibinfo{year}{2019}),
  \bibinfo{pages}{505--510}.
\newblock
\showISSN{1476-4687}
\urldef\tempurl%
\url{https://doi.org/10.1038/s41586-019-1666-5}
\showURL{%
\tempurl}


\bibitem[{Ball} et~al\mbox{.}(2021)]%
        {ball2021software}
\bibfield{author}{\bibinfo{person}{Harrison {Ball}},
  \bibinfo{person}{Michael~J. {Biercuk}}, \bibinfo{person}{Andre R.~R.
  {Carvalho}}, \bibinfo{person}{Jiayin {Chen}}, \bibinfo{person}{Michael
  {Hush}}, \bibinfo{person}{Leonardo~A. {De Castro}}, \bibinfo{person}{Li
  {Li}}, \bibinfo{person}{Per~J. {Liebermann}}, \bibinfo{person}{Harry~J.
  {Slatyer}}, \bibinfo{person}{Claire {Edmunds}}, \bibinfo{person}{Virginia
  {Frey}}, \bibinfo{person}{Cornelius {Hempel}}, {and}
  \bibinfo{person}{Alistair {Milne}}.} \bibinfo{year}{2021}\natexlab{}.
\newblock \showarticletitle{Software tools for quantum control: Improving
  quantum computer performance through noise and error suppression}.
\newblock \bibinfo{journal}{\emph{Quantum Science and Technology}}
  \bibinfo{volume}{6}, \bibinfo{number}{4} (\bibinfo{year}{2021}),
  \bibinfo{pages}{044011}.
\newblock
\urldef\tempurl%
\url{https://doi.org/10.1088/2058-9565/abdca6}
\showURL{%
\tempurl}


\bibitem[Beverland et~al\mbox{.}(2021)]%
        {beverland2021cost}
\bibfield{author}{\bibinfo{person}{Michael~E Beverland},
  \bibinfo{person}{Aleksander Kubica}, {and} \bibinfo{person}{Krysta~M Svore}.}
  \bibinfo{year}{2021}\natexlab{}.
\newblock \showarticletitle{Cost of universality: A comparative study of the
  overhead of state distillation and code switching with color codes}.
\newblock \bibinfo{journal}{\emph{PRX Quantum}} \bibinfo{volume}{2},
  \bibinfo{number}{2} (\bibinfo{year}{2021}), \bibinfo{pages}{020341}.
\newblock
\urldef\tempurl%
\url{https://doi.org/10.1103/PRXQuantum.2.020341}
\showURL{%
\tempurl}


\bibitem[Beverland et~al\mbox{.}(2022)]%
        {beverland2022assessing}
\bibfield{author}{\bibinfo{person}{Michael~E Beverland},
  \bibinfo{person}{Prakash Murali}, \bibinfo{person}{Matthias Troyer},
  \bibinfo{person}{Krysta~M Svore}, \bibinfo{person}{Torsten Hoefler},
  \bibinfo{person}{Vadym Kliuchnikov}, \bibinfo{person}{Guang~Hao Low},
  \bibinfo{person}{Mathias Soeken}, \bibinfo{person}{Aarthi Sundaram}, {and}
  \bibinfo{person}{Alexander Vaschillo}.} \bibinfo{year}{2022}\natexlab{}.
\newblock \showarticletitle{Assessing requirements to scale to practical
  quantum advantage}.
\newblock \bibinfo{journal}{\emph{arXiv preprint arXiv:2211.07629}}
  (\bibinfo{year}{2022}).
\newblock
\urldef\tempurl%
\url{https://doi.org/10.48550/arXiv.2211.07629}
\showURL{%
\tempurl}


\bibitem[Bravyi and Haah(2012)]%
        {bravyi2012magic}
\bibfield{author}{\bibinfo{person}{Sergey Bravyi} {and}
  \bibinfo{person}{Jeongwan Haah}.} \bibinfo{year}{2012}\natexlab{}.
\newblock \showarticletitle{Magic-state distillation with low overhead}.
\newblock \bibinfo{journal}{\emph{Physical Review A—Atomic, Molecular, and
  Optical Physics}} \bibinfo{volume}{86}, \bibinfo{number}{5}
  (\bibinfo{year}{2012}), \bibinfo{pages}{052329}.
\newblock
\urldef\tempurl%
\url{https://doi.org/10.1103/PhysRevA.86.052329}
\showURL{%
\tempurl}


\bibitem[Bravyi and Kitaev(2005)]%
        {bravyi2005universal}
\bibfield{author}{\bibinfo{person}{Sergey Bravyi} {and} \bibinfo{person}{Alexei
  Kitaev}.} \bibinfo{year}{2005}\natexlab{}.
\newblock \showarticletitle{Universal quantum computation with ideal Clifford
  gates and noisy ancillas}.
\newblock \bibinfo{journal}{\emph{Physical Review A—Atomic, Molecular, and
  Optical Physics}} \bibinfo{volume}{71}, \bibinfo{number}{2}
  (\bibinfo{year}{2005}), \bibinfo{pages}{022316}.
\newblock
\urldef\tempurl%
\url{https://doi.org/10.1103/PhysRevA.71.022316}
\showURL{%
\tempurl}


\bibitem[Buluta and Nori(2009)]%
        {buluta2009quantum}
\bibfield{author}{\bibinfo{person}{Iulia Buluta} {and} \bibinfo{person}{Franco
  Nori}.} \bibinfo{year}{2009}\natexlab{}.
\newblock \showarticletitle{Quantum simulators}.
\newblock \bibinfo{journal}{\emph{Science}} \bibinfo{volume}{326},
  \bibinfo{number}{5949} (\bibinfo{year}{2009}), \bibinfo{pages}{108--111}.
\newblock
\urldef\tempurl%
\url{https://doi.org/10.1126/science.1177838}
\showURL{%
\tempurl}


\bibitem[Butt et~al\mbox{.}(2024)]%
        {butt2024fault}
\bibfield{author}{\bibinfo{person}{Friederike Butt}, \bibinfo{person}{Sascha
  Heu{\ss}en}, \bibinfo{person}{Manuel Rispler}, {and} \bibinfo{person}{Markus
  M{\"u}ller}.} \bibinfo{year}{2024}\natexlab{}.
\newblock \showarticletitle{Fault-tolerant code-switching protocols for
  near-term quantum processors}.
\newblock \bibinfo{journal}{\emph{PRX Quantum}} \bibinfo{volume}{5},
  \bibinfo{number}{2} (\bibinfo{year}{2024}), \bibinfo{pages}{020345}.
\newblock
\urldef\tempurl%
\url{https://doi.org/10.1103/PRXQuantum.5.020345}
\showURL{%
\tempurl}


\bibitem[Campbell(2014)]%
        {campbell2014enhanced}
\bibfield{author}{\bibinfo{person}{Earl~T Campbell}.}
  \bibinfo{year}{2014}\natexlab{}.
\newblock \showarticletitle{Enhanced fault-tolerant quantum computing in
  d-level systems}.
\newblock \bibinfo{journal}{\emph{Physical review letters}}
  \bibinfo{volume}{113}, \bibinfo{number}{23} (\bibinfo{year}{2014}),
  \bibinfo{pages}{230501}.
\newblock
\urldef\tempurl%
\url{https://doi.org/10.1103/PhysRevLett.113.230501}
\showURL{%
\tempurl}


\bibitem[Campbell et~al\mbox{.}(2012)]%
        {campbell2012magic}
\bibfield{author}{\bibinfo{person}{Earl~T Campbell}, \bibinfo{person}{Hussain
  Anwar}, {and} \bibinfo{person}{Dan~E Browne}.}
  \bibinfo{year}{2012}\natexlab{}.
\newblock \showarticletitle{Magic-state distillation in all prime dimensions
  using quantum reed-muller codes}.
\newblock \bibinfo{journal}{\emph{Physical Review X}} \bibinfo{volume}{2},
  \bibinfo{number}{4} (\bibinfo{year}{2012}), \bibinfo{pages}{041021}.
\newblock
\urldef\tempurl%
\url{https://doi.org/10.1103/PhysRevX.2.041021}
\showURL{%
\tempurl}


\bibitem[Carrera~Vazquez et~al\mbox{.}(2024)]%
        {carrera2024combining}
\bibfield{author}{\bibinfo{person}{Almudena Carrera~Vazquez},
  \bibinfo{person}{Caroline Tornow}, \bibinfo{person}{Diego Rist{\`e}},
  \bibinfo{person}{Stefan Woerner}, \bibinfo{person}{Maika Takita}, {and}
  \bibinfo{person}{Daniel~J Egger}.} \bibinfo{year}{2024}\natexlab{}.
\newblock \showarticletitle{Combining quantum processors with real-time
  classical communication}.
\newblock \bibinfo{journal}{\emph{Nature}} (\bibinfo{year}{2024}),
  \bibinfo{pages}{1--5}.
\newblock
\urldef\tempurl%
\url{https://doi.org/10.1038/s41586-024-08178-2}
\showURL{%
\tempurl}


\bibitem[Chamberland et~al\mbox{.}(2022)]%
        {chamberland2022building}
\bibfield{author}{\bibinfo{person}{Christopher Chamberland},
  \bibinfo{person}{Kyungjoo Noh}, \bibinfo{person}{Patricio Arrangoiz-Arriola},
  \bibinfo{person}{Earl~T. Campbell}, \bibinfo{person}{Connor~T. Hann},
  \bibinfo{person}{Joseph Iverson}, \bibinfo{person}{Harald Putterman},
  \bibinfo{person}{Thomas~C. Bohdanowicz}, \bibinfo{person}{Steven~T. Flammia},
  \bibinfo{person}{Andrew Keller}, \bibinfo{person}{Gil Refael},
  \bibinfo{person}{John Preskill}, \bibinfo{person}{Liang Jiang},
  \bibinfo{person}{Amir~H. Safavi-Naeini}, \bibinfo{person}{Oskar Painter},
  {and} \bibinfo{person}{Fernando~G.S.L. Brand\~ao}.}
  \bibinfo{year}{2022}\natexlab{}.
\newblock \showarticletitle{Building a fault-tolerant quantum computer using
  concatenated cat codes}.
\newblock \bibinfo{journal}{\emph{PRX Quantum}} \bibinfo{volume}{3},
  \bibinfo{number}{1} (\bibinfo{year}{2022}), \bibinfo{pages}{010329}.
\newblock
\urldef\tempurl%
\url{https://doi.org/10.1103/PRXQuantum.3.010329}
\showURL{%
\tempurl}


\bibitem[Chatterjee et~al\mbox{.}(2025)]%
        {chatterjee2025q}
\bibfield{author}{\bibinfo{person}{Avimita Chatterjee},
  \bibinfo{person}{Archisman Ghosh}, {and} \bibinfo{person}{Swaroop Ghosh}.}
  \bibinfo{year}{2025}\natexlab{}.
\newblock \showarticletitle{The Q-Spellbook: Crafting Surface Code Layouts and
  Magic State Protocols for Large-Scale Quantum Computing}.
\newblock \bibinfo{journal}{\emph{arXiv preprint arXiv:2502.11253}}
  (\bibinfo{year}{2025}).
\newblock
\urldef\tempurl%
\url{https://doi.org/10.48550/arXiv.2502.11253}
\showURL{%
\tempurl}


\bibitem[Choi et~al\mbox{.}(2023)]%
        {choi2023fault}
\bibfield{author}{\bibinfo{person}{Hyeongrak Choi}, \bibinfo{person}{Frederic~T
  Chong}, \bibinfo{person}{Dirk Englund}, {and} \bibinfo{person}{Yongshan
  Ding}.} \bibinfo{year}{2023}\natexlab{}.
\newblock \showarticletitle{Fault tolerant non-clifford state preparation for
  arbitrary rotations}.
\newblock \bibinfo{journal}{\emph{arXiv preprint arXiv:2303.17380}}
  (\bibinfo{year}{2023}).
\newblock
\urldef\tempurl%
\url{https://doi.org/10.48550/arXiv.2303.17380}
\showURL{%
\tempurl}


\bibitem[Dadpour et~al\mbox{.}(2025)]%
        {dadpour2025low}
\bibfield{author}{\bibinfo{person}{Amir~H Dadpour}, \bibinfo{person}{Timur
  Khayrullin}, \bibinfo{person}{Fouad Afiouni}, \bibinfo{person}{Remy~El
  Sabeh}, \bibinfo{person}{Amer~E Mouawad}, \bibinfo{person}{Izzat~El Hajj},
  {and} \bibinfo{person}{Alexandre Cooper}.} \bibinfo{year}{2025}\natexlab{}.
\newblock \showarticletitle{Low-latency control system for feedback experiments
  with optical tweezer arrays}.
\newblock \bibinfo{journal}{\emph{arXiv preprint arXiv:2504.06528}}
  (\bibinfo{year}{2025}).
\newblock
\urldef\tempurl%
\url{https://doi.org/10.48550/arXiv.2504.06528}
\showURL{%
\tempurl}


\bibitem[Demeulemeester and Herroelen(2002)]%
        {demeulemeester2002project}
\bibfield{author}{\bibinfo{person}{Erik~L Demeulemeester} {and}
  \bibinfo{person}{Willy~S Herroelen}.} \bibinfo{year}{2002}\natexlab{}.
\newblock \bibinfo{booktitle}{\emph{Project scheduling: a research handbook}}.
\newblock \bibinfo{publisher}{Springer}.
\newblock
\urldef\tempurl%
\url{https://doi.org/10.1007/b101924}
\showURL{%
\tempurl}


\bibitem[Devoret and Schoelkopf(2013)]%
        {devoret2013superconducting}
\bibfield{author}{\bibinfo{person}{Michel~H Devoret} {and}
  \bibinfo{person}{Robert~J Schoelkopf}.} \bibinfo{year}{2013}\natexlab{}.
\newblock \showarticletitle{Superconducting circuits for quantum information:
  an outlook}.
\newblock \bibinfo{journal}{\emph{Science}} \bibinfo{volume}{339},
  \bibinfo{number}{6124} (\bibinfo{year}{2013}), \bibinfo{pages}{1169--1174}.
\newblock
\urldef\tempurl%
\url{https://doi.org/10.1126/science.1231930}
\showURL{%
\tempurl}


\bibitem[Ding et~al\mbox{.}(2024)]%
        {ding2024experimental}
\bibfield{author}{\bibinfo{person}{Chunyang Ding}, \bibinfo{person}{Martin
  Di~Federico}, \bibinfo{person}{Michael Hatridge}, \bibinfo{person}{Andrew
  Houck}, \bibinfo{person}{Sebastien Leger}, \bibinfo{person}{Jeronimo
  Martinez}, \bibinfo{person}{Connie Miao}, \bibinfo{person}{David~Schuster I},
  \bibinfo{person}{Leandro Stefanazzi}, \bibinfo{person}{Chris Stoughton},
  \bibinfo{person}{Sara Sussman}, \bibinfo{person}{Ken Treptow},
  \bibinfo{person}{Sho Uemura}, \bibinfo{person}{Neal Wilcer},
  \bibinfo{person}{Helin Zhang}, \bibinfo{person}{Chao Zhou}, {and}
  \bibinfo{person}{Gustavo Cancelo}.} \bibinfo{year}{2024}\natexlab{}.
\newblock \showarticletitle{Experimental advances with the QICK (Quantum
  Instrumentation Control Kit) for superconducting quantum hardware}.
\newblock \bibinfo{journal}{\emph{Physical Review Research}}
  \bibinfo{volume}{6}, \bibinfo{number}{1} (\bibinfo{year}{2024}),
  \bibinfo{pages}{013305}.
\newblock
\urldef\tempurl%
\url{https://doi.org/10.1103/PhysRevResearch.6.013305}
\showURL{%
\tempurl}


\bibitem[Ding et~al\mbox{.}(2018)]%
        {ding2018magic}
\bibfield{author}{\bibinfo{person}{Yongshan Ding}, \bibinfo{person}{Adam
  Holmes}, \bibinfo{person}{Ali Javadi-Abhari}, \bibinfo{person}{Diana
  Franklin}, \bibinfo{person}{Margaret Martonosi}, {and}
  \bibinfo{person}{Frederic Chong}.} \bibinfo{year}{2018}\natexlab{}.
\newblock \showarticletitle{Magic-state functional units: Mapping and
  scheduling multi-level distillation circuits for fault-tolerant quantum
  architectures}. In \bibinfo{booktitle}{\emph{2018 51st Annual IEEE/ACM
  International Symposium on Microarchitecture (MICRO)}}. IEEE,
  \bibinfo{pages}{828--840}.
\newblock
\urldef\tempurl%
\url{https://doi.org/10.1109/MICRO.2018.00072}
\showURL{%
\tempurl}


\bibitem[Feynman(1982)]%
        {feynman2018simulating}
\bibfield{author}{\bibinfo{person}{Richard~P Feynman}.}
  \bibinfo{year}{1982}\natexlab{}.
\newblock \showarticletitle{Simulating physics with computers}.
\newblock \bibinfo{journal}{\emph{International Journal of Theoretical
  Physics}} \bibinfo{volume}{21}, \bibinfo{number}{6} (\bibinfo{date}{June}
  \bibinfo{year}{1982}), \bibinfo{pages}{467--488}.
\newblock
\urldef\tempurl%
\url{https://doi.org/10.1007/BF02650179}
\showURL{%
\tempurl}


\bibitem[Fowler et~al\mbox{.}(2013)]%
        {fowler2013surface}
\bibfield{author}{\bibinfo{person}{Austin~G Fowler}, \bibinfo{person}{Simon~J
  Devitt}, {and} \bibinfo{person}{Cody Jones}.}
  \bibinfo{year}{2013}\natexlab{}.
\newblock \showarticletitle{Surface code implementation of block code state
  distillation}.
\newblock \bibinfo{journal}{\emph{Scientific reports}} \bibinfo{volume}{3},
  \bibinfo{number}{1} (\bibinfo{year}{2013}), \bibinfo{pages}{1939}.
\newblock
\urldef\tempurl%
\url{https://doi.org/10.1038/srep01939}
\showURL{%
\tempurl}


\bibitem[Fowler and Gidney(2018)]%
        {fowler2018low}
\bibfield{author}{\bibinfo{person}{Austin~G Fowler} {and}
  \bibinfo{person}{Craig Gidney}.} \bibinfo{year}{2018}\natexlab{}.
\newblock \showarticletitle{Low overhead quantum computation using lattice
  surgery}.
\newblock \bibinfo{journal}{\emph{arXiv preprint arXiv:1808.06709}}
  (\bibinfo{year}{2018}).
\newblock
\urldef\tempurl%
\url{https://doi.org/10.48550/arXiv.1808.06709}
\showURL{%
\tempurl}


\bibitem[Gidney(2023)]%
        {gidney2023cleaner}
\bibfield{author}{\bibinfo{person}{Craig Gidney}.}
  \bibinfo{year}{2023}\natexlab{}.
\newblock \showarticletitle{Cleaner magic states with hook injection}.
\newblock \bibinfo{journal}{\emph{arXiv preprint arXiv:2302.12292}}
  (\bibinfo{year}{2023}).
\newblock
\urldef\tempurl%
\url{https://doi.org/10.48550/arXiv.2302.12292}
\showURL{%
\tempurl}


\bibitem[Gidney and Eker{\aa}(2021)]%
        {gidney2021factor}
\bibfield{author}{\bibinfo{person}{Craig Gidney} {and} \bibinfo{person}{Martin
  Eker{\aa}}.} \bibinfo{year}{2021}\natexlab{}.
\newblock \showarticletitle{How to factor 2048 bit RSA integers in 8 hours
  using 20 million noisy qubits}.
\newblock \bibinfo{journal}{\emph{Quantum}}  \bibinfo{volume}{5}
  (\bibinfo{year}{2021}), \bibinfo{pages}{433}.
\newblock
\urldef\tempurl%
\url{https://doi.org/10.22331/q-2021-04-15-433}
\showURL{%
\tempurl}


\bibitem[Gidney and Fowler(2019)]%
        {gidney2019efficient}
\bibfield{author}{\bibinfo{person}{Craig Gidney} {and}
  \bibinfo{person}{Austin~G Fowler}.} \bibinfo{year}{2019}\natexlab{}.
\newblock \showarticletitle{Efficient magic state factories with a catalyzed $|
  CCZ\rangle $ to $2| T\rangle $ transformation}.
\newblock \bibinfo{journal}{\emph{Quantum}}  \bibinfo{volume}{3}
  (\bibinfo{year}{2019}), \bibinfo{pages}{135}.
\newblock
\urldef\tempurl%
\url{https://doi.org/10.22331/q-2019-04-30-135}
\showURL{%
\tempurl}


\bibitem[Gidney et~al\mbox{.}(2024)]%
        {gidney2024cultivation}
\bibfield{author}{\bibinfo{person}{Craig Gidney}, \bibinfo{person}{Noah
  Shutty}, {and} \bibinfo{person}{Cody Jones}.}
  \bibinfo{year}{2024}\natexlab{}.
\newblock \showarticletitle{Magic state cultivation: growing T states as cheap
  as CNOT gates}.
\newblock \bibinfo{journal}{\emph{arXiv preprint arXiv:2409.17595}}
  (\bibinfo{year}{2024}).
\newblock
\urldef\tempurl%
\url{https://doi.org/10.48550/arXiv.2409.17595}
\showURL{%
\tempurl}


\bibitem[Giortamis et~al\mbox{.}(2024)]%
        {giortamis2024orchestrating}
\bibfield{author}{\bibinfo{person}{Emmanouil Giortamis},
  \bibinfo{person}{Francisco Rom{\~a}o}, \bibinfo{person}{Nathaniel Tornow},
  \bibinfo{person}{Dmitry Lugovoy}, {and} \bibinfo{person}{Pramod Bhatotia}.}
  \bibinfo{year}{2024}\natexlab{}.
\newblock \showarticletitle{Orchestrating quantum cloud environments with
  qonductor}.
\newblock \bibinfo{journal}{\emph{arXiv preprint arXiv:2408.04312}}
  (\bibinfo{year}{2024}).
\newblock
\urldef\tempurl%
\url{https://doi.org/10.48550/arXiv.2408.04312}
\showURL{%
\tempurl}


\bibitem[Gottesman(1998)]%
        {gottesman1998heisenberg}
\bibfield{author}{\bibinfo{person}{Daniel Gottesman}.}
  \bibinfo{year}{1998}\natexlab{}.
\newblock \showarticletitle{The Heisenberg representation of quantum
  computers}.
\newblock \bibinfo{journal}{\emph{arXiv preprint quant-ph/9807006}}
  (\bibinfo{year}{1998}).
\newblock
\urldef\tempurl%
\url{https://doi.org/10.48550/arXiv.quant-ph/9807006}
\showURL{%
\tempurl}


\bibitem[{Gurobi Optimization, LLC}(2024)]%
        {gurobi}
\bibfield{author}{\bibinfo{person}{{Gurobi Optimization, LLC}}.}
  \bibinfo{year}{2024}\natexlab{}.
\newblock \bibinfo{title}{{Gurobi Optimizer Reference Manual}}.
\newblock
\urldef\tempurl%
\url{https://www.gurobi.com}
\showURL{%
\tempurl}


\bibitem[Haah and Hastings(2018)]%
        {haah2018codes}
\bibfield{author}{\bibinfo{person}{Jeongwan Haah} {and}
  \bibinfo{person}{Matthew~B Hastings}.} \bibinfo{year}{2018}\natexlab{}.
\newblock \showarticletitle{Codes and protocols for distilling $ t $,
  controlled-$ s $, and toffoli gates}.
\newblock \bibinfo{journal}{\emph{Quantum}}  \bibinfo{volume}{2}
  (\bibinfo{year}{2018}), \bibinfo{pages}{71}.
\newblock
\urldef\tempurl%
\url{https://doi.org/10.22331/q-2018-06-07-71}
\showURL{%
\tempurl}


\bibitem[Haah et~al\mbox{.}(2017a)]%
        {haah2017magicintermediate}
\bibfield{author}{\bibinfo{person}{Jeongwan Haah}, \bibinfo{person}{Matthew~B
  Hastings}, \bibinfo{person}{David Poulin}, {and} \bibinfo{person}{Dave
  Wecker}.} \bibinfo{year}{2017}\natexlab{a}.
\newblock \showarticletitle{Magic state distillation at intermediate size}.
\newblock \bibinfo{journal}{\emph{arXiv preprint arXiv:1709.02789}}
  (\bibinfo{year}{2017}).
\newblock
\urldef\tempurl%
\url{https://doi.org/10.48550/arXiv.1709.02789}
\showURL{%
\tempurl}


\bibitem[Haah et~al\mbox{.}(2017b)]%
        {haah2017magic}
\bibfield{author}{\bibinfo{person}{Jeongwan Haah}, \bibinfo{person}{Matthew~B
  Hastings}, \bibinfo{person}{David Poulin}, {and} \bibinfo{person}{Dave
  Wecker}.} \bibinfo{year}{2017}\natexlab{b}.
\newblock \showarticletitle{Magic state distillation with low space overhead
  and optimal asymptotic input count}.
\newblock \bibinfo{journal}{\emph{Quantum}}  \bibinfo{volume}{1}
  (\bibinfo{year}{2017}), \bibinfo{pages}{31}.
\newblock
\urldef\tempurl%
\url{https://doi.org/10.22331/q-2017-10-03-31}
\showURL{%
\tempurl}


\bibitem[Harrigan et~al\mbox{.}(2024)]%
        {harrigan2024qualtran}
\bibfield{author}{\bibinfo{person}{Matthew~P. Harrigan}, \bibinfo{person}{Tanuj
  Khattar}, \bibinfo{person}{Charles Yuan}, \bibinfo{person}{Anurudh Peduri},
  \bibinfo{person}{Noureldin Yosri}, \bibinfo{person}{Fionn~D. Malone},
  \bibinfo{person}{Ryan Babbush}, {and} \bibinfo{person}{Nicholas~C. Rubin}.}
  \bibinfo{year}{2024}\natexlab{}.
\newblock \bibinfo{title}{Expressing and Analyzing Quantum Algorithms with
  Qualtran}.
\newblock
\showeprint[arxiv]{2409.04643}~[quant-ph]
\urldef\tempurl%
\url{https://doi.org/10.48550/arXiv.2409.04643}
\showURL{%
\tempurl}


\bibitem[Hastings and Haah(2018)]%
        {hastings2018distillation}
\bibfield{author}{\bibinfo{person}{Matthew~B Hastings} {and}
  \bibinfo{person}{Jeongwan Haah}.} \bibinfo{year}{2018}\natexlab{}.
\newblock \showarticletitle{Distillation with sublogarithmic overhead}.
\newblock \bibinfo{journal}{\emph{Physical review letters}}
  \bibinfo{volume}{120}, \bibinfo{number}{5} (\bibinfo{year}{2018}),
  \bibinfo{pages}{050504}.
\newblock
\urldef\tempurl%
\url{https://doi.org/10.1103/PhysRevLett.120.050504}
\showURL{%
\tempurl}


\bibitem[Hatano and Suzuki(2005)]%
        {hatano2005finding}
\bibfield{author}{\bibinfo{person}{Naomichi Hatano} {and}
  \bibinfo{person}{Masuo Suzuki}.} \bibinfo{year}{2005}\natexlab{}.
\newblock \showarticletitle{Finding exponential product formulas of higher
  orders}.
\newblock In \bibinfo{booktitle}{\emph{Quantum annealing and other optimization
  methods}}. \bibinfo{publisher}{Springer}, \bibinfo{pages}{37--68}.
\newblock
\urldef\tempurl%
\url{https://doi.org/10.1007/11526216_2}
\showURL{%
\tempurl}


\bibitem[Heu{\ss}en and Hilder(2024)]%
        {heussen2024efficient}
\bibfield{author}{\bibinfo{person}{Sascha Heu{\ss}en} {and}
  \bibinfo{person}{Janine Hilder}.} \bibinfo{year}{2024}\natexlab{}.
\newblock \showarticletitle{Efficient fault-tolerant code switching via one-way
  transversal CNOT gates}.
\newblock \bibinfo{journal}{\emph{arXiv preprint arXiv:2409.13465}}
  (\bibinfo{year}{2024}).
\newblock
\urldef\tempurl%
\url{https://doi.org/10.48550/arXiv.2409.13465}
\showURL{%
\tempurl}


\bibitem[Hietala et~al\mbox{.}(2021)]%
        {hietala2021verified}
\bibfield{author}{\bibinfo{person}{Kesha Hietala}, \bibinfo{person}{Robert
  Rand}, \bibinfo{person}{Shih-Han Hung}, \bibinfo{person}{Xiaodi Wu}, {and}
  \bibinfo{person}{Michael Hicks}.} \bibinfo{year}{2021}\natexlab{}.
\newblock \showarticletitle{A verified optimizer for quantum circuits}.
\newblock \bibinfo{journal}{\emph{Proceedings of the ACM on Programming
  Languages}} \bibinfo{volume}{5}, \bibinfo{number}{POPL}
  (\bibinfo{year}{2021}), \bibinfo{pages}{1--29}.
\newblock
\urldef\tempurl%
\url{https://doi.org/10.1145/3434318}
\showURL{%
\tempurl}


\bibitem[Hirano and Fujii(2025)]%
        {hirano2025locality}
\bibfield{author}{\bibinfo{person}{Yutaka Hirano} {and}
  \bibinfo{person}{Keisuke Fujii}.} \bibinfo{year}{2025}\natexlab{}.
\newblock \showarticletitle{Locality-aware Pauli-based computation for local
  magic state preparation}.
\newblock \bibinfo{journal}{\emph{arXiv preprint arXiv:2504.12091}}
  (\bibinfo{year}{2025}).
\newblock
\urldef\tempurl%
\url{https://doi.org/10.48550/arXiv.2504.12091}
\showURL{%
\tempurl}


\bibitem[Hirano et~al\mbox{.}(2024a)]%
        {hirano2024leveraging}
\bibfield{author}{\bibinfo{person}{Yutaka Hirano}, \bibinfo{person}{Tomohiro
  Itogawa}, {and} \bibinfo{person}{Keisuke Fujii}.}
  \bibinfo{year}{2024}\natexlab{a}.
\newblock \showarticletitle{Leveraging zero-level distillation to generate
  high-fidelity magic states}. In \bibinfo{booktitle}{\emph{2024 IEEE
  International Conference on Quantum Computing and Engineering (QCE)}},
  Vol.~\bibinfo{volume}{1}. IEEE, \bibinfo{pages}{843--853}.
\newblock
\urldef\tempurl%
\url{https://doi.org/10.1109/QCE60285.2024.00104}
\showURL{%
\tempurl}


\bibitem[Hirano et~al\mbox{.}(2024b)]%
        {hirano2024magicpool}
\bibfield{author}{\bibinfo{person}{Yutaka Hirano}, \bibinfo{person}{Yasunari
  Suzuki}, {and} \bibinfo{person}{Keisuke Fujii}.}
  \bibinfo{year}{2024}\natexlab{b}.
\newblock \showarticletitle{Magicpool: Dealing with magic state distillation
  failures on large-scale fault-tolerant quantum computer}.
\newblock \bibinfo{journal}{\emph{arXiv preprint arXiv:2407.07394}}
  (\bibinfo{year}{2024}).
\newblock
\urldef\tempurl%
\url{https://doi.org/10.48550/arXiv.2407.07394}
\showURL{%
\tempurl}


\bibitem[Hoefler et~al\mbox{.}(2023)]%
        {hoefler2023disentangling}
\bibfield{author}{\bibinfo{person}{Torsten Hoefler}, \bibinfo{person}{Thomas
  H{\"a}ner}, {and} \bibinfo{person}{Matthias Troyer}.}
  \bibinfo{year}{2023}\natexlab{}.
\newblock \showarticletitle{Disentangling hype from practicality: On
  realistically achieving quantum advantage}.
\newblock \bibinfo{journal}{\emph{Commun. ACM}} \bibinfo{volume}{66},
  \bibinfo{number}{5} (\bibinfo{year}{2023}), \bibinfo{pages}{82--87}.
\newblock
\urldef\tempurl%
\url{https://doi.org/10.1145/3571725}
\showURL{%
\tempurl}


\bibitem[Holmes et~al\mbox{.}(2019)]%
        {holmes2019resource}
\bibfield{author}{\bibinfo{person}{Adam Holmes}, \bibinfo{person}{Yongshan
  Ding}, \bibinfo{person}{Ali Javadi-Abhari}, \bibinfo{person}{Diana Franklin},
  \bibinfo{person}{Margaret Martonosi}, {and} \bibinfo{person}{Frederic~T
  Chong}.} \bibinfo{year}{2019}\natexlab{}.
\newblock \showarticletitle{Resource optimized quantum architectures for
  surface code implementations of magic-state distillation}.
\newblock \bibinfo{journal}{\emph{Microprocessors and Microsystems}}
  \bibinfo{volume}{67} (\bibinfo{year}{2019}), \bibinfo{pages}{56--70}.
\newblock
\urldef\tempurl%
\url{https://doi.org/10.1016/j.micpro.2019.02.007}
\showURL{%
\tempurl}


\bibitem[Hornibrook et~al\mbox{.}(2015)]%
        {hornibrook2015cryogenic}
\bibfield{author}{\bibinfo{person}{J.~M. Hornibrook}, \bibinfo{person}{J.~I.
  Colless}, \bibinfo{person}{I.~D. Conway~Lamb}, \bibinfo{person}{S.~J. Pauka},
  \bibinfo{person}{H. Lu}, \bibinfo{person}{A.~C. Gossard},
  \bibinfo{person}{J.~D. Watson}, \bibinfo{person}{G.~C. Gardner},
  \bibinfo{person}{S. Fallahi}, \bibinfo{person}{M.~J. Manfra}, {and}
  \bibinfo{person}{D.~J. Reilly}.} \bibinfo{year}{2015}\natexlab{}.
\newblock \showarticletitle{Cryogenic control architecture for large-scale
  quantum computing}.
\newblock \bibinfo{journal}{\emph{Physical Review Applied}}
  \bibinfo{volume}{3}, \bibinfo{number}{2} (\bibinfo{year}{2015}),
  \bibinfo{pages}{024010}.
\newblock
\urldef\tempurl%
\url{https://doi.org/10.1103/PhysRevApplied.3.024010}
\showURL{%
\tempurl}


\bibitem[Horsman et~al\mbox{.}(2012)]%
        {horsman2012surface}
\bibfield{author}{\bibinfo{person}{Dominic Horsman}, \bibinfo{person}{Austin~G
  Fowler}, \bibinfo{person}{Simon Devitt}, {and} \bibinfo{person}{Rodney
  Van~Meter}.} \bibinfo{year}{2012}\natexlab{}.
\newblock \showarticletitle{Surface code quantum computing by lattice surgery}.
\newblock \bibinfo{journal}{\emph{New Journal of Physics}}
  \bibinfo{volume}{14}, \bibinfo{number}{12} (\bibinfo{year}{2012}),
  \bibinfo{pages}{123011}.
\newblock
\urldef\tempurl%
\url{https://doi.org/10.1088/1367-2630/14/12/123011}
\showURL{%
\tempurl}


\bibitem[Hua et~al\mbox{.}(2023)]%
        {hua2023caqr}
\bibfield{author}{\bibinfo{person}{Fei Hua}, \bibinfo{person}{Yuwei Jin},
  \bibinfo{person}{Yanhao Chen}, \bibinfo{person}{Suhas Vittal},
  \bibinfo{person}{Kevin Krsulich}, \bibinfo{person}{Lev~S Bishop},
  \bibinfo{person}{John Lapeyre}, \bibinfo{person}{Ali Javadi-Abhari}, {and}
  \bibinfo{person}{Eddy~Z Zhang}.} \bibinfo{year}{2023}\natexlab{}.
\newblock \showarticletitle{Caqr: A compiler-assisted approach for qubit reuse
  through dynamic circuit}. In \bibinfo{booktitle}{\emph{Proceedings of the
  28th ACM International Conference on Architectural Support for Programming
  Languages and Operating Systems, Volume 3}}. \bibinfo{pages}{59--71}.
\newblock
\urldef\tempurl%
\url{https://doi.org/10.1145/3582016.3582030}
\showURL{%
\tempurl}


\bibitem[Ishizaka and Nemery(2013)]%
        {ishizaka2013multi}
\bibfield{author}{\bibinfo{person}{Alessio Ishizaka} {and}
  \bibinfo{person}{Philippe Nemery}.} \bibinfo{year}{2013}\natexlab{}.
\newblock \bibinfo{booktitle}{\emph{Multi-criteria decision analysis: methods
  and software}}.
\newblock \bibinfo{publisher}{John Wiley \& Sons}.
\newblock
\urldef\tempurl%
\url{https://doi.org/10.1007/978-1-4757-3157-6_2}
\showURL{%
\tempurl}


\bibitem[Itogawa et~al\mbox{.}(2024)]%
        {itogawa2024even}
\bibfield{author}{\bibinfo{person}{Tomohiro Itogawa}, \bibinfo{person}{Yugo
  Takada}, \bibinfo{person}{Yutaka Hirano}, {and} \bibinfo{person}{Keisuke
  Fujii}.} \bibinfo{year}{2024}\natexlab{}.
\newblock \showarticletitle{Even more efficient magic state distillation by
  zero-level distillation}.
\newblock \bibinfo{journal}{\emph{arXiv preprint arXiv:2403.03991}}
  (\bibinfo{year}{2024}).
\newblock
\urldef\tempurl%
\url{https://doi.org/10.48550/arXiv.2403.03991}
\showURL{%
\tempurl}


\bibitem[Jiang(2024)]%
        {jiang2024qubit}
\bibfield{author}{\bibinfo{person}{Hanru Jiang}.}
  \bibinfo{year}{2024}\natexlab{}.
\newblock \showarticletitle{Qubit recycling revisited}.
\newblock \bibinfo{journal}{\emph{Proceedings of the ACM on Programming
  Languages}} \bibinfo{volume}{8}, \bibinfo{number}{PLDI}
  (\bibinfo{year}{2024}), \bibinfo{pages}{1264--1287}.
\newblock
\urldef\tempurl%
\url{https://doi.org/10.1145/3656428}
\showURL{%
\tempurl}


\bibitem[Jones(2013)]%
        {jones2013multilevel}
\bibfield{author}{\bibinfo{person}{Cody Jones}.}
  \bibinfo{year}{2013}\natexlab{}.
\newblock \showarticletitle{Multilevel distillation of magic states for quantum
  computing}.
\newblock \bibinfo{journal}{\emph{Physical Review A—Atomic, Molecular, and
  Optical Physics}} \bibinfo{volume}{87}, \bibinfo{number}{4}
  (\bibinfo{year}{2013}), \bibinfo{pages}{042305}.
\newblock
\urldef\tempurl%
\url{https://doi.org/10.1103/PhysRevA.87.042305}
\showURL{%
\tempurl}


\bibitem[{Jurcevic} et~al\mbox{.}(2021)]%
        {jurcevic2021demonstration}
\bibfield{author}{\bibinfo{person}{Petar {Jurcevic}}, \bibinfo{person}{Ali
  {Javadi-Abhari}}, \bibinfo{person}{Lev~S. {Bishop}}, \bibinfo{person}{Isaac
  {Lauer}}, \bibinfo{person}{Daniela~F. {Bogorin}}, \bibinfo{person}{Markus
  {Brink}}, \bibinfo{person}{Lauren {Capelluto}}, \bibinfo{person}{Oktay
  {G{\"u}nl{\"u}k}}, \bibinfo{person}{Toshinari {Itoko}},
  \bibinfo{person}{Naoki {Kanazawa}}, \bibinfo{person}{Abhinav {Kandala}},
  \bibinfo{person}{George~A. {Keefe}}, \bibinfo{person}{Kevin {Krsulich}},
  \bibinfo{person}{William {Landers}}, \bibinfo{person}{Eric~P. {Lewandowski}},
  \bibinfo{person}{Douglas~T. {McClure}}, \bibinfo{person}{Giacomo
  {Nannicini}}, \bibinfo{person}{Adinath {Narasgond}},
  \bibinfo{person}{Hasan~M. {Nayfeh}}, \bibinfo{person}{Emily {Pritchett}},
  \bibinfo{person}{Mary~Beth {Rothwell}}, \bibinfo{person}{Srikanth
  {Srinivasan}}, \bibinfo{person}{Neereja {Sundaresan}}, \bibinfo{person}{Cindy
  {Wang}}, \bibinfo{person}{Ken~X. {Wei}}, \bibinfo{person}{Christopher~J.
  {Wood}}, \bibinfo{person}{Jeng-Bang {Yau}}, \bibinfo{person}{Eric~J.
  {Zhang}}, \bibinfo{person}{Oliver~E. {Dial}}, \bibinfo{person}{Jerry~M.
  {Chow}}, {and} \bibinfo{person}{Jay~M. {Gambetta}}.}
  \bibinfo{year}{2021}\natexlab{}.
\newblock \showarticletitle{Demonstration of quantum volume 64 on a
  superconducting quantum computing system}.
\newblock \bibinfo{journal}{\emph{Quantum Science and Technology}}
  \bibinfo{volume}{6}, \bibinfo{number}{2} (\bibinfo{year}{2021}),
  \bibinfo{pages}{025020}.
\newblock
\urldef\tempurl%
\url{https://doi.org/10.1088/2058-9565/abe519}
\showURL{%
\tempurl}


\bibitem[Kan et~al\mbox{.}(2025)]%
        {kan2025sparo}
\bibfield{author}{\bibinfo{person}{Shuwen Kan}, \bibinfo{person}{Zefan Du},
  \bibinfo{person}{Chenxu Liu}, \bibinfo{person}{Meng Wang},
  \bibinfo{person}{Yufei Ding}, \bibinfo{person}{Ang Li}, \bibinfo{person}{Ying
  Mao}, {and} \bibinfo{person}{Samuel Stein}.} \bibinfo{year}{2025}\natexlab{}.
\newblock \showarticletitle{SPARO: Surface-code Pauli-based Architectural
  Resource Optimization for Fault-tolerant Quantum Computing}.
\newblock \bibinfo{journal}{\emph{arXiv preprint arXiv:2504.21854}}
  (\bibinfo{year}{2025}).
\newblock
\urldef\tempurl%
\url{https://doi.org/10.48550/arXiv.2504.21854}
\showURL{%
\tempurl}


\bibitem[Karzig et~al\mbox{.}(2017)]%
        {karzig2017scalable}
\bibfield{author}{\bibinfo{person}{Torsten Karzig}, \bibinfo{person}{Christina
  Knapp}, \bibinfo{person}{Roman~M. Lutchyn}, \bibinfo{person}{Parsa
  Bonderson}, \bibinfo{person}{Matthew~B. Hastings}, \bibinfo{person}{Chetan
  Nayak}, \bibinfo{person}{Jason Alicea}, \bibinfo{person}{Karsten Flensberg},
  \bibinfo{person}{Stephan Plugge}, \bibinfo{person}{Yuval Oreg},
  \bibinfo{person}{Charles~M. Marcus}, {and} \bibinfo{person}{Michael~H.
  Freedman}.} \bibinfo{year}{2017}\natexlab{}.
\newblock \showarticletitle{Scalable designs for
  quasiparticle-poisoning-protected topological quantum computation with
  Majorana zero modes}.
\newblock \bibinfo{journal}{\emph{Phys. Rev. B}}  \bibinfo{volume}{95}
  (\bibinfo{date}{Jun} \bibinfo{year}{2017}), \bibinfo{pages}{235305}.
\newblock
Issue 23.
\urldef\tempurl%
\url{https://doi.org/10.1103/PhysRevB.95.235305}
\showURL{%
\tempurl}


\bibitem[Kissinger and Van De~Wetering(2019)]%
        {kissinger2019pyzx}
\bibfield{author}{\bibinfo{person}{Aleks Kissinger} {and} \bibinfo{person}{John
  Van De~Wetering}.} \bibinfo{year}{2019}\natexlab{}.
\newblock \showarticletitle{PyZX: Large scale automated diagrammatic
  reasoning}.
\newblock \bibinfo{journal}{\emph{arXiv preprint arXiv:1904.04735}}
  (\bibinfo{year}{2019}).
\newblock
\urldef\tempurl%
\url{https://doi.org/10.48550/arXiv.1904.04735}
\showURL{%
\tempurl}


\bibitem[Kjaergaard et~al\mbox{.}(2020)]%
        {kjaergaard2020superconducting}
\bibfield{author}{\bibinfo{person}{Morten Kjaergaard},
  \bibinfo{person}{Mollie~E Schwartz}, \bibinfo{person}{Jochen Braum{\"u}ller},
  \bibinfo{person}{Philip Krantz}, \bibinfo{person}{Joel I-J Wang},
  \bibinfo{person}{Simon Gustavsson}, {and} \bibinfo{person}{William~D
  Oliver}.} \bibinfo{year}{2020}\natexlab{}.
\newblock \showarticletitle{Superconducting qubits: Current state of play}.
\newblock \bibinfo{journal}{\emph{Annual Review of Condensed Matter Physics}}
  \bibinfo{volume}{11}, \bibinfo{number}{1} (\bibinfo{year}{2020}),
  \bibinfo{pages}{369--395}.
\newblock
\urldef\tempurl%
\url{https://doi.org/10.1146/annurev-conmatphys-031119-050605}
\showURL{%
\tempurl}


\bibitem[Knill(2004)]%
        {knill2004fault}
\bibfield{author}{\bibinfo{person}{Emanuel Knill}.}
  \bibinfo{year}{2004}\natexlab{}.
\newblock \showarticletitle{Fault-tolerant postselected quantum computation:
  Schemes}.
\newblock \bibinfo{journal}{\emph{arXiv preprint quant-ph/0402171}}
  (\bibinfo{year}{2004}).
\newblock
\urldef\tempurl%
\url{https://doi.org/10.48550/arXiv.quant-ph/0402171}
\showURL{%
\tempurl}


\bibitem[Lao and Criger(2022)]%
        {lao2022magic}
\bibfield{author}{\bibinfo{person}{Lingling Lao} {and} \bibinfo{person}{Ben
  Criger}.} \bibinfo{year}{2022}\natexlab{}.
\newblock \showarticletitle{Magic state injection on the rotated surface code}.
  In \bibinfo{booktitle}{\emph{Proceedings of the 19th ACM International
  Conference on Computing Frontiers}}. \bibinfo{pages}{113--120}.
\newblock
\urldef\tempurl%
\url{https://doi.org/10.1145/3528416.3530237}
\showURL{%
\tempurl}


\bibitem[Lee et~al\mbox{.}(2025)]%
        {lee2025low}
\bibfield{author}{\bibinfo{person}{Seok-Hyung Lee}, \bibinfo{person}{Felix
  Thomsen}, \bibinfo{person}{Nicholas Fazio}, \bibinfo{person}{Benjamin~J
  Brown}, {and} \bibinfo{person}{Stephen~D Bartlett}.}
  \bibinfo{year}{2025}\natexlab{}.
\newblock \showarticletitle{Low-overhead magic state distillation with color
  codes}.
\newblock \bibinfo{journal}{\emph{PRX Quantum}} \bibinfo{volume}{6},
  \bibinfo{number}{3} (\bibinfo{year}{2025}), \bibinfo{pages}{030317}.
\newblock
\urldef\tempurl%
\url{https://doi.org/10.1103/ch5r-cnfq}
\showURL{%
\tempurl}


\bibitem[Li(2015)]%
        {li2015magic}
\bibfield{author}{\bibinfo{person}{Ying Li}.} \bibinfo{year}{2015}\natexlab{}.
\newblock \showarticletitle{A magic state’s fidelity can be superior to the
  operations that created it}.
\newblock \bibinfo{journal}{\emph{New Journal of Physics}}
  \bibinfo{volume}{17}, \bibinfo{number}{2} (\bibinfo{year}{2015}),
  \bibinfo{pages}{023037}.
\newblock
\urldef\tempurl%
\url{https://doi.org/10.1088/1367-2630/17/2/023037}
\showURL{%
\tempurl}


\bibitem[Litinski(2019a)]%
        {litinski2019game}
\bibfield{author}{\bibinfo{person}{Daniel Litinski}.}
  \bibinfo{year}{2019}\natexlab{a}.
\newblock \showarticletitle{A game of surface codes: Large-scale quantum
  computing with lattice surgery}.
\newblock \bibinfo{journal}{\emph{Quantum}}  \bibinfo{volume}{3}
  (\bibinfo{year}{2019}), \bibinfo{pages}{128}.
\newblock
\urldef\tempurl%
\url{https://doi.org/10.22331/q-2019-03-05-128}
\showURL{%
\tempurl}


\bibitem[Litinski(2019b)]%
        {litinski2019magic}
\bibfield{author}{\bibinfo{person}{Daniel Litinski}.}
  \bibinfo{year}{2019}\natexlab{b}.
\newblock \showarticletitle{Magic state distillation: Not as costly as you
  think}.
\newblock \bibinfo{journal}{\emph{Quantum}}  \bibinfo{volume}{3}
  (\bibinfo{year}{2019}), \bibinfo{pages}{205}.
\newblock
\urldef\tempurl%
\url{https://doi.org/10.22331/q-2019-12-02-205}
\showURL{%
\tempurl}


\bibitem[Lloyd(1996)]%
        {lloyd1996universal}
\bibfield{author}{\bibinfo{person}{Seth Lloyd}.}
  \bibinfo{year}{1996}\natexlab{}.
\newblock \showarticletitle{Universal quantum simulators}.
\newblock \bibinfo{journal}{\emph{Science}} \bibinfo{volume}{273},
  \bibinfo{number}{5278} (\bibinfo{year}{1996}), \bibinfo{pages}{1073--1078}.
\newblock
\urldef\tempurl%
\url{https://doi.org/10.1126/science.273.5278.1073}
\showURL{%
\tempurl}


\bibitem[Low and Chuang(2017)]%
        {low2017optimal}
\bibfield{author}{\bibinfo{person}{Guang~Hao Low} {and}
  \bibinfo{person}{Isaac~L Chuang}.} \bibinfo{year}{2017}\natexlab{}.
\newblock \showarticletitle{Optimal Hamiltonian simulation by quantum signal
  processing}.
\newblock \bibinfo{journal}{\emph{Physical review letters}}
  \bibinfo{volume}{118}, \bibinfo{number}{1} (\bibinfo{year}{2017}),
  \bibinfo{pages}{010501}.
\newblock
\urldef\tempurl%
\url{https://doi.org/10.1103/PhysRevLett.118.010501}
\showURL{%
\tempurl}


\bibitem[Meier et~al\mbox{.}(2012)]%
        {meier2012magic}
\bibfield{author}{\bibinfo{person}{Adam~M Meier}, \bibinfo{person}{Bryan
  Eastin}, {and} \bibinfo{person}{Emanuel Knill}.}
  \bibinfo{year}{2012}\natexlab{}.
\newblock \showarticletitle{Magic-state distillation with the four-qubit code}.
\newblock \bibinfo{journal}{\emph{arXiv preprint arXiv:1204.4221}}
  (\bibinfo{year}{2012}).
\newblock
\urldef\tempurl%
\url{https://doi.org/10.48550/arXiv.1204.4221}
\showURL{%
\tempurl}


\bibitem[Mermin(2007)]%
        {Mermin_2007}
\bibfield{author}{\bibinfo{person}{N.~David Mermin}.}
  \bibinfo{year}{2007}\natexlab{}.
\newblock \bibinfo{booktitle}{\emph{Quantum Computer Science: An
  Introduction}}.
\newblock \bibinfo{publisher}{Cambridge University Press}.
\newblock
\urldef\tempurl%
\url{https://doi.org/10.1017/CBO9780511813870}
\showURL{%
\tempurl}


\bibitem[{Microsoft}(2023)]%
        {Microsoft_Azure_Quantum_Development}
\bibfield{author}{\bibinfo{person}{{Microsoft}}.}
  \bibinfo{year}{2023}\natexlab{}.
\newblock \bibinfo{booktitle}{\emph{{Azure Quantum Development Kit}}}.
\newblock
\urldef\tempurl%
\url{https://github.com/microsoft/qsharp}
\showURL{%
\tempurl}
\newblock
\shownote{"Accessed: 2025-08-10"}.


\bibitem[Paler et~al\mbox{.}(2016)]%
        {paler2016wire}
\bibfield{author}{\bibinfo{person}{Alexandru Paler}, \bibinfo{person}{Robert
  Wille}, {and} \bibinfo{person}{Simon~J Devitt}.}
  \bibinfo{year}{2016}\natexlab{}.
\newblock \showarticletitle{Wire recycling for quantum circuit optimization}.
\newblock \bibinfo{journal}{\emph{Physical Review A}} \bibinfo{volume}{94},
  \bibinfo{number}{4} (\bibinfo{year}{2016}), \bibinfo{pages}{042337}.
\newblock
\urldef\tempurl%
\url{https://doi.org/10.1103/PhysRevA.94.042337}
\showURL{%
\tempurl}


\bibitem[Patel et~al\mbox{.}(2022)]%
        {patel2022quest}
\bibfield{author}{\bibinfo{person}{Tirthak Patel}, \bibinfo{person}{Ed Younis},
  \bibinfo{person}{Costin Iancu}, \bibinfo{person}{Wibe de Jong}, {and}
  \bibinfo{person}{Devesh Tiwari}.} \bibinfo{year}{2022}\natexlab{}.
\newblock \showarticletitle{QUEST: systematically approximating Quantum
  circuits for higher output fidelity}. In
  \bibinfo{booktitle}{\emph{Proceedings of the 27th ACM International
  Conference on Architectural Support for Programming Languages and Operating
  Systems}} (Lausanne, Switzerland) \emph{(\bibinfo{series}{ASPLOS '22})}.
  \bibinfo{publisher}{Association for Computing Machinery},
  \bibinfo{address}{New York, NY, USA}, \bibinfo{pages}{514–528}.
\newblock
\showISBNx{9781450392051}
\urldef\tempurl%
\url{https://doi.org/10.1145/3503222.3507739}
\showURL{%
\tempurl}


\bibitem[Pogorelov et~al\mbox{.}(2025)]%
        {pogorelov2025experimental}
\bibfield{author}{\bibinfo{person}{Ivan Pogorelov}, \bibinfo{person}{Friederike
  Butt}, \bibinfo{person}{Lukas Postler}, \bibinfo{person}{Christian~D
  Marciniak}, \bibinfo{person}{Philipp Schindler}, \bibinfo{person}{Markus
  M{\"u}ller}, {and} \bibinfo{person}{Thomas Monz}.}
  \bibinfo{year}{2025}\natexlab{}.
\newblock \showarticletitle{Experimental fault-tolerant code switching}.
\newblock \bibinfo{journal}{\emph{Nature Physics}} \bibinfo{volume}{21},
  \bibinfo{number}{2} (\bibinfo{year}{2025}), \bibinfo{pages}{298--303}.
\newblock
\urldef\tempurl%
\url{https://doi.org/10.1038/s41567-024-02727-2}
\showURL{%
\tempurl}


\bibitem[Prabhu and Chamberland(2022)]%
        {prabhu2022new}
\bibfield{author}{\bibinfo{person}{Prithviraj Prabhu} {and}
  \bibinfo{person}{Christopher Chamberland}.} \bibinfo{year}{2022}\natexlab{}.
\newblock \showarticletitle{New magic state distillation factories optimized by
  temporally encoded lattice surgery}.
\newblock \bibinfo{journal}{\emph{arXiv preprint arXiv:2210.15814}}
  (\bibinfo{year}{2022}).
\newblock
\urldef\tempurl%
\url{https://doi.org/10.48550/arXiv.2210.15814}
\showURL{%
\tempurl}


\bibitem[Preskill(2018)]%
        {Preskill2018nisq}
\bibfield{author}{\bibinfo{person}{John Preskill}.}
  \bibinfo{year}{2018}\natexlab{}.
\newblock \showarticletitle{Quantum Computing in the NISQ era and beyond}.
\newblock \bibinfo{journal}{\emph{{Quantum}}}  \bibinfo{volume}{2}
  (\bibinfo{date}{Aug.} \bibinfo{year}{2018}), \bibinfo{pages}{79}.
\newblock
\showISSN{2521-327X}
\urldef\tempurl%
\url{https://doi.org/10.22331/q-2018-08-06-79}
\showURL{%
\tempurl}


\bibitem[Preskill(2023)]%
        {preskill2023quantum}
\bibfield{author}{\bibinfo{person}{John Preskill}.}
  \bibinfo{year}{2023}\natexlab{}.
\newblock \showarticletitle{Quantum computing 40 years later}.
\newblock In \bibinfo{booktitle}{\emph{Feynman Lectures on Computation}}.
  \bibinfo{publisher}{CRC Press}, \bibinfo{pages}{193--244}.
\newblock
\urldef\tempurl%
\url{https://doi.org/10.48550/arXiv.2106.10522}
\showURL{%
\tempurl}


\bibitem[Qin et~al\mbox{.}(2019)]%
        {qin2019fpga}
\bibfield{author}{\bibinfo{person}{Xi Qin}, \bibinfo{person}{Wenzhe Zhang},
  \bibinfo{person}{Lin Wang}, \bibinfo{person}{Yuxi Zhao}, \bibinfo{person}{Yu
  Tong}, \bibinfo{person}{Xing Rong}, {and} \bibinfo{person}{Jiangfeng Du}.}
  \bibinfo{year}{2019}\natexlab{}.
\newblock \showarticletitle{An FPGA-based hardware platform for the control of
  spin-based quantum systems}.
\newblock \bibinfo{journal}{\emph{IEEE Transactions on Instrumentation and
  Measurement}} \bibinfo{volume}{69}, \bibinfo{number}{4}
  (\bibinfo{year}{2019}), \bibinfo{pages}{1127--1139}.
\newblock
\urldef\tempurl%
\url{https://doi.org/10.1109/TIM.2019.2910921}
\showURL{%
\tempurl}


\bibitem[Sethi and Baker(2025)]%
        {sethi2025rescq}
\bibfield{author}{\bibinfo{person}{Sayam Sethi} {and}
  \bibinfo{person}{Jonathan~Mark Baker}.} \bibinfo{year}{2025}\natexlab{}.
\newblock \showarticletitle{RESCQ: Realtime Scheduling for Continuous Angle
  Quantum Error Correction Architectures}. In
  \bibinfo{booktitle}{\emph{Proceedings of the 30th ACM International
  Conference on Architectural Support for Programming Languages and Operating
  Systems, Volume 2}} (Rotterdam, Netherlands) \emph{(\bibinfo{series}{ASPLOS
  '25})}. \bibinfo{publisher}{Association for Computing Machinery},
  \bibinfo{address}{New York, NY, USA}, \bibinfo{pages}{1028–1043}.
\newblock
\showISBNx{9798400710797}
\urldef\tempurl%
\url{https://doi.org/10.1145/3676641.3716018}
\showURL{%
\tempurl}


\bibitem[Shor(1999)]%
        {shor1999polynomial}
\bibfield{author}{\bibinfo{person}{Peter~W Shor}.}
  \bibinfo{year}{1999}\natexlab{}.
\newblock \showarticletitle{Polynomial-time algorithms for prime factorization
  and discrete logarithms on a quantum computer}.
\newblock \bibinfo{journal}{\emph{SIAM review}} \bibinfo{volume}{41},
  \bibinfo{number}{2} (\bibinfo{year}{1999}), \bibinfo{pages}{303--332}.
\newblock
\urldef\tempurl%
\url{https://doi.org/10.1137/S0036144598347011}
\showURL{%
\tempurl}


\bibitem[{Silva} et~al\mbox{.}(2024)]%
        {silva2024optimizing}
\bibfield{author}{\bibinfo{person}{Allyson {Silva}}, \bibinfo{person}{Artur
  {Scherer}}, \bibinfo{person}{Zak {Webb}}, \bibinfo{person}{Abdullah
  {Khalid}}, \bibinfo{person}{Bohdan {Kulchytskyy}}, \bibinfo{person}{Mia
  {Kramer}}, \bibinfo{person}{Kevin {Nguyen}}, \bibinfo{person}{Xiangzhou
  {Kong}}, \bibinfo{person}{Gebremedhin~A. {Dagnew}}, \bibinfo{person}{Yumeng
  {Wang}}, \bibinfo{person}{Huy~Anh {Nguyen}}, \bibinfo{person}{Einar
  {Gabbassov}}, \bibinfo{person}{Katiemarie {Olfert}}, {and}
  \bibinfo{person}{Pooya {Ronagh}}.} \bibinfo{year}{2024}\natexlab{}.
\newblock \showarticletitle{Optimizing multi-level magic state factories for
  fault-tolerant quantum architectures}.
\newblock \bibinfo{journal}{\emph{arXiv preprint arXiv:2411.04270}}
  (\bibinfo{year}{2024}).
\newblock
\urldef\tempurl%
\url{https://doi.org/10.48550/arXiv.2411.04270}
\showURL{%
\tempurl}


\bibitem[Silva et~al\mbox{.}(2024)]%
        {silva2024multi}
\bibfield{author}{\bibinfo{person}{Allyson Silva}, \bibinfo{person}{Xiangyi
  Zhang}, \bibinfo{person}{Zak Webb}, \bibinfo{person}{Mia Kramer},
  \bibinfo{person}{Chan~Woo Yang}, \bibinfo{person}{Xiao Liu},
  \bibinfo{person}{Jessica Lemieux}, \bibinfo{person}{Ka-Wai Chen},
  \bibinfo{person}{Artur Scherer}, {and} \bibinfo{person}{Pooya Ronagh}.}
  \bibinfo{year}{2024}\natexlab{}.
\newblock \showarticletitle{Multi-qubit lattice surgery scheduling}.
\newblock \bibinfo{journal}{\emph{arXiv preprint arXiv:2405.17688}}
  (\bibinfo{year}{2024}).
\newblock
\urldef\tempurl%
\url{https://doi.org/10.48550/arXiv.2405.17688}
\showURL{%
\tempurl}


\bibitem[Singh et~al\mbox{.}(2022)]%
        {singh2022high}
\bibfield{author}{\bibinfo{person}{Shraddha Singh}, \bibinfo{person}{Andrew~S
  Darmawan}, \bibinfo{person}{Benjamin~J Brown}, {and} \bibinfo{person}{Shruti
  Puri}.} \bibinfo{year}{2022}\natexlab{}.
\newblock \showarticletitle{High-fidelity magic-state preparation with a
  biased-noise architecture}.
\newblock \bibinfo{journal}{\emph{Physical Review A}} \bibinfo{volume}{105},
  \bibinfo{number}{5} (\bibinfo{year}{2022}), \bibinfo{pages}{052410}.
\newblock
\urldef\tempurl%
\url{https://doi.org/10.1103/PhysRevA.105.052410}
\showURL{%
\tempurl}


\bibitem[Stanco et~al\mbox{.}(2022)]%
        {stanco2022versatile}
\bibfield{author}{\bibinfo{person}{Andrea Stanco},
  \bibinfo{person}{Francesco~BL Santagiustina}, \bibinfo{person}{Luca
  Calderaro}, \bibinfo{person}{Marco Avesani}, \bibinfo{person}{Tommaso
  Bertapelle}, \bibinfo{person}{Daniele Dequal}, \bibinfo{person}{Giuseppe
  Vallone}, {and} \bibinfo{person}{Paolo Villoresi}.}
  \bibinfo{year}{2022}\natexlab{}.
\newblock \showarticletitle{Versatile and concurrent FPGA-based architecture
  for practical quantum communication systems}.
\newblock \bibinfo{journal}{\emph{IEEE Transactions on Quantum Engineering}}
  \bibinfo{volume}{3} (\bibinfo{year}{2022}), \bibinfo{pages}{1--8}.
\newblock
\urldef\tempurl%
\url{https://doi.org/10.1109/TQE.2022.3143997}
\showURL{%
\tempurl}


\bibitem[Stefanazzi et~al\mbox{.}(2022)]%
        {stefanazzi2022qick}
\bibfield{author}{\bibinfo{person}{Leandro Stefanazzi},
  \bibinfo{person}{Kenneth Treptow}, \bibinfo{person}{Neal Wilcer},
  \bibinfo{person}{Chris Stoughton}, \bibinfo{person}{Collin Bradford},
  \bibinfo{person}{Sho Uemura}, \bibinfo{person}{Silvia Zorzetti},
  \bibinfo{person}{Salvatore Montella}, \bibinfo{person}{Gustavo Cancelo},
  \bibinfo{person}{Sara Sussman}, \bibinfo{person}{Andrew Houck},
  \bibinfo{person}{Shefali Saxena}, \bibinfo{person}{Horacio Arnaldi},
  \bibinfo{person}{Ankur Agrawal}, \bibinfo{person}{Helin Zhang},
  \bibinfo{person}{Chunyang Ding}, {and} \bibinfo{person}{David~I. Schuster}.}
  \bibinfo{year}{2022}\natexlab{}.
\newblock \showarticletitle{The QICK (Quantum Instrumentation Control Kit):
  Readout and control for qubits and detectors}.
\newblock \bibinfo{journal}{\emph{Review of Scientific Instruments}}
  \bibinfo{volume}{93}, \bibinfo{number}{4} (\bibinfo{year}{2022}).
\newblock
\urldef\tempurl%
\url{https://doi.org/10.1063/5.0076249}
\showURL{%
\tempurl}


\bibitem[Tan et~al\mbox{.}(2024)]%
        {tan2024sat}
\bibfield{author}{\bibinfo{person}{Daniel~Bochen Tan},
  \bibinfo{person}{Murphy~Yuezhen Niu}, {and} \bibinfo{person}{Craig Gidney}.}
  \bibinfo{year}{2024}\natexlab{}.
\newblock \showarticletitle{A SAT Scalpel for Lattice Surgery: Representation
  and Synthesis of Subroutines for Surface-Code Fault-Tolerant Quantum
  Computing}. In \bibinfo{booktitle}{\emph{2024 ACM/IEEE 51st Annual
  International Symposium on Computer Architecture (ISCA)}}. IEEE,
  \bibinfo{pages}{325--339}.
\newblock
\urldef\tempurl%
\url{https://doi.org/10.1109/ISCA59077.2024.00032}
\showURL{%
\tempurl}


\bibitem[Vaknin et~al\mbox{.}(2025)]%
        {vaknin2025magic}
\bibfield{author}{\bibinfo{person}{Yotam Vaknin}, \bibinfo{person}{Shoham
  Jacoby}, \bibinfo{person}{Arne Grimsmo}, {and} \bibinfo{person}{Alex
  Retzker}.} \bibinfo{year}{2025}\natexlab{}.
\newblock \showarticletitle{Magic State Cultivation on the Surface Code}.
\newblock \bibinfo{journal}{\emph{arXiv preprint arXiv:2502.01743}}
  (\bibinfo{year}{2025}).
\newblock
\urldef\tempurl%
\url{https://doi.org/10.48550/arXiv.2502.01743}
\showURL{%
\tempurl}


\bibitem[van Dam et~al\mbox{.}(2023)]%
        {Azure_Quantum_Resource_Estimator}
\bibfield{author}{\bibinfo{person}{Wim van Dam}, \bibinfo{person}{Mariia
  Mykhailova}, {and} \bibinfo{person}{Mathias Soeken}.}
  \bibinfo{year}{2023}\natexlab{}.
\newblock \showarticletitle{{Using Azure Quantum Resource Estimator for
  Assessing Performance of Fault Tolerant Quantum Computation}}. In
  \bibinfo{booktitle}{\emph{Proceedings of the SC '23 Workshops of The
  International Conference on High Performance Computing, Network, Storage, and
  Analysis}} \emph{(\bibinfo{series}{SC-W '23})}.
  \bibinfo{publisher}{Association for Computing Machinery},
  \bibinfo{address}{New York, NY, USA}, \bibinfo{pages}{1414–1419}.
\newblock
\showISBNx{9798400707858}
\urldef\tempurl%
\url{https://doi.org/10.1145/3624062.3624211}
\showURL{%
\tempurl}


\bibitem[von Burg et~al\mbox{.}(2021)]%
        {von2021quantum}
\bibfield{author}{\bibinfo{person}{Vera von Burg}, \bibinfo{person}{Guang~Hao
  Low}, \bibinfo{person}{Thomas H{\"a}ner}, \bibinfo{person}{Damian~S Steiger},
  \bibinfo{person}{Markus Reiher}, \bibinfo{person}{Martin Roetteler}, {and}
  \bibinfo{person}{Matthias Troyer}.} \bibinfo{year}{2021}\natexlab{}.
\newblock \showarticletitle{Quantum computing enhanced computational
  catalysis}.
\newblock \bibinfo{journal}{\emph{Physical Review Research}}
  \bibinfo{volume}{3}, \bibinfo{number}{3} (\bibinfo{year}{2021}),
  \bibinfo{pages}{033055}.
\newblock
\urldef\tempurl%
\url{https://doi.org/10.1103/PhysRevResearch.3.033055}
\showURL{%
\tempurl}


\bibitem[Wan(2024)]%
        {wan2024constant}
\bibfield{author}{\bibinfo{person}{Kwok~Ho Wan}.}
  \bibinfo{year}{2024}\natexlab{}.
\newblock \showarticletitle{Constant-time magic state distillation}.
\newblock \bibinfo{journal}{\emph{arXiv preprint arXiv:2410.17992}}
  (\bibinfo{year}{2024}).
\newblock
\urldef\tempurl%
\url{https://doi.org/10.48550/arXiv.2410.17992}
\showURL{%
\tempurl}


\bibitem[Wan et~al\mbox{.}(2024)]%
        {wan2024iterative}
\bibfield{author}{\bibinfo{person}{Kwok~Ho Wan}, \bibinfo{person}{Mark Webber},
  \bibinfo{person}{Austin~G Fowler}, {and} \bibinfo{person}{Winfried~K
  Hensinger}.} \bibinfo{year}{2024}\natexlab{}.
\newblock \showarticletitle{An iterative transversal CNOT decoder}.
\newblock \bibinfo{journal}{\emph{arXiv preprint arXiv:2407.20976}}
  (\bibinfo{year}{2024}).
\newblock
\urldef\tempurl%
\url{https://doi.org/10.48550/arXiv.2407.20976}
\showURL{%
\tempurl}


\bibitem[Xu et~al\mbox{.}(2023)]%
        {xu2023synthesizing}
\bibfield{author}{\bibinfo{person}{Amanda Xu}, \bibinfo{person}{Abtin Molavi},
  \bibinfo{person}{Lauren Pick}, \bibinfo{person}{Swamit Tannu}, {and}
  \bibinfo{person}{Aws Albarghouthi}.} \bibinfo{year}{2023}\natexlab{}.
\newblock \showarticletitle{Synthesizing quantum-circuit optimizers}.
\newblock \bibinfo{journal}{\emph{Proceedings of the ACM on Programming
  Languages}} \bibinfo{volume}{7}, \bibinfo{number}{PLDI}
  (\bibinfo{year}{2023}), \bibinfo{pages}{835--859}.
\newblock
\urldef\tempurl%
\url{https://doi.org/10.1145/3591254}
\showURL{%
\tempurl}


\bibitem[Xu et~al\mbox{.}(2025)]%
        {xu2025optimizing}
\bibfield{author}{\bibinfo{person}{Amanda Xu}, \bibinfo{person}{Abtin Molavi},
  \bibinfo{person}{Swamit Tannu}, {and} \bibinfo{person}{Aws Albarghouthi}.}
  \bibinfo{year}{2025}\natexlab{}.
\newblock \showarticletitle{Optimizing quantum circuits, fast and slow}. In
  \bibinfo{booktitle}{\emph{Proceedings of the 30th ACM International
  Conference on Architectural Support for Programming Languages and Operating
  Systems, Volume 1}}. \bibinfo{pages}{777--793}.
\newblock
\urldef\tempurl%
\url{https://doi.org/10.1145/3669940.3707240}
\showURL{%
\tempurl}


\bibitem[Xu et~al\mbox{.}(2021)]%
        {xu2021qubic}
\bibfield{author}{\bibinfo{person}{Yilun Xu}, \bibinfo{person}{Gang Huang},
  \bibinfo{person}{Jan Balewski}, \bibinfo{person}{Ravi Naik},
  \bibinfo{person}{Alexis Morvan}, \bibinfo{person}{Bradley Mitchell},
  \bibinfo{person}{Kasra Nowrouzi}, \bibinfo{person}{David~I Santiago}, {and}
  \bibinfo{person}{Irfan Siddiqi}.} \bibinfo{year}{2021}\natexlab{}.
\newblock \showarticletitle{QubiC: An open-source FPGA-based control and
  measurement system for superconducting quantum information processors}.
\newblock \bibinfo{journal}{\emph{IEEE Transactions on Quantum Engineering}}
  \bibinfo{volume}{2} (\bibinfo{year}{2021}), \bibinfo{pages}{1--11}.
\newblock
\urldef\tempurl%
\url{https://doi.org/10.1109/TQE.2021.3116540}
\showURL{%
\tempurl}


\bibitem[Zhu et~al\mbox{.}(2019)]%
        {zhu2019training}
\bibfield{author}{\bibinfo{person}{D. Zhu}, \bibinfo{person}{N.~M. Linke},
  \bibinfo{person}{M. Benedetti}, \bibinfo{person}{K.~A. Landsman},
  \bibinfo{person}{N.~H. Nguyen}, \bibinfo{person}{C.~H. Alderete},
  \bibinfo{person}{A. Perdomo-Ortiz}, \bibinfo{person}{N. Korda},
  \bibinfo{person}{A. Garfoot}, \bibinfo{person}{C. Brecque},
  \bibinfo{person}{L. Egan}, \bibinfo{person}{O. Perdomo}, {and}
  \bibinfo{person}{C. Monroe}.} \bibinfo{year}{2019}\natexlab{}.
\newblock \showarticletitle{Training of quantum circuits on a hybrid quantum
  computer}.
\newblock \bibinfo{journal}{\emph{Science advances}} \bibinfo{volume}{5},
  \bibinfo{number}{10} (\bibinfo{year}{2019}), \bibinfo{pages}{eaaw9918}.
\newblock
\urldef\tempurl%
\url{https://doi.org/10.1126/sciadv.aaw9918}
\showURL{%
\tempurl}


\end{thebibliography}

\appendix
\section{Calculate expected delays due to failures}
\label{sec:appendix}

In this section, we analytically estimate the expected delays caused by low-level factory failures.
For simplicity, we assume that each factory outputs only one magic state at a time.
However, this can be easily extended to the general case.

We first run the no-failure simulation according to Section~\ref{sec:setup} to obtain the series of numbers of magic states produced $n_\text{prod}(t)$ and consumed $n_\text{cons}(t)$ at each time $t$.
We observe that this is a Markov process and model it with state vector
\begin{equation}
\mathbf{P}_t = \Big( p_t(0), p_t(1), \dots, p_t(N_\text{buf}),p_t(\text{fail}) \Big),
\end{equation}
where $p_t(i)$ represents the probability that the buffer contains exactly $i$ magic states at time $t$, and $p_t(\text{fail})$ represents the probability that the system has encountered its first stall due to insufficient magic states by time $t$.
State evolution is given by
\begin{equation}
\mathbf{P}_t = \mathbf{P}_{t-1} \cdot \mathbf{T}_\text{cons} \cdot \mathbf{T}_\text{prod},
\end{equation}
where $\mathbf{T}_\text{cons}$ and $\mathbf{T}_\text{prod}$ are the transition matrices for consumption and production, respectively.

When the high-level factory consumes $n_\text{cons}(t)$ magic states, the probability vector shifts by $n_\text{cons}(t)$ positions, as consumption is deterministic.
If the buffer contains fewer than $n_\text{cons}(t)$ magic states, a stall occurs.
Thus,
\begin{equation}
\mathbf{T}_\text{cons}(i, j) =
\begin{cases}
1, & \text{if } j = i - n_\text{cons}(t), \\
0, & \text{otherwise}.
\end{cases}
\end{equation}
The last element $p_t(\text{fail})$ accumulates the probability of a stall as
\begin{equation}
p_t(\text{fail}) = p_{t-1}(\text{fail}) + \sum_{i=0}^{n_\text{cons}(t)-1} p_{t-1}(i).
\end{equation}
On the production side, the number of magic states generated by low-level factories follows a binomial distribution, updating the probability vector as
\begin{equation}
\mathbf{T}_\text{prod}(i, j) = \binom{n_\text{prod}(t)}{j-i} p_\text{suc}^{j-i} (1 - p_\text{suc})^{n_\text{prod}(t)-(j-i)},
\end{equation}
where $p_\text{suc}$ is the success probability for producing a single magic state.

In this analytical model, the probability of a stall $P_\text{stall}(t)$ at each time step $t$ is actually the cumulative probability that at least one stall has occurred up to and including time $t$.
Therefore, the cumulative stall probability is $P_\text{stall}(t) = p_t(\text{fail})$.
The probability that the first stall occurs at time $t$ should be computed as
\begin{equation}
p_\text{stall}(t) = P_\text{stall}(t) - P_\text{stall}(t-1),
\end{equation}
where $P_\text{stall}(-1)$ is defined as $0$.

To compute the expected failure delay, we weight the recovery time $\Delta t(t)$ at each $t$ by the stall probability and get the expected delay
\begin{equation}
    \mathbb{E}[T_\text{delay}] = \sum_{t=0}^T p_\text{stall}(t) \cdot \Delta t(t),
\end{equation}
where $T$ is the total time of the distillation process.
To get the $\Delta t(t)$, we simply use the integer programming algorithm as described in Section~\ref{sec:5:spatial}.
This estimated delay $\mathbb{E}[T_\text{delay}]$ is added to the total time of the distillation factory.

In this approach, we only account for the delay associated with the first occurrence of a stall that is caused by low-level factory failures.
After recovery from this stall, the system could encounter additional stalls due to further low-level factory failures.
However, the probability of such subsequent stalls is extremely low so it can be safely neglected for practical results.

\end{document}